\documentclass[acmsmall,screen]{acmart}%
\setcopyright{rightsretained}
\acmYear{2026}
\copyrightyear{2026}
\acmVolume{3}
\acmArticle{1}
\acmMonth{10}

\newtheorem{remark}{Remark}
\newcommand{\ToolName}{\textsc{Opt}}

\usepackage{threeparttable}
\usepackage[inline]{enumitem}
\usepackage{url}
\usepackage{amsthm, newtxmath}

\newcommand{\mytodoblue}[1]{\textcolor{blue}{\ding{46}~{\sf}~#1}}
\newcommand{\mytodored}[1]{\textcolor{red}{\ding{46}~{\sf}~#1}}

\newcommand{\mytodocyan}[1]{\textcolor{cyan}{\ding{46}~{\sf}~#1}}

\usepackage{pgfplots} % violin plot
\pgfplotsset{compat=1.18}
\usepgfplotslibrary{statistics}

\usepackage{booktabs}
\usepackage{color,xcolor}
\usepackage{mathtools}
\usepackage{multirow}
\usepackage{tcolorbox}
\usepackage{listings}
\usepackage{courier}
\usepackage{cancel}
\usepackage[ruled,linesnumbered]{algorithm2e}

\usepackage{pifont}
\usepackage{subcaption}
\usepackage[utf8]{inputenc}
\usepackage{cleveref}
\crefname{section}{§}{§§}
\Crefname{section}{§}{§§}
\usepackage{tcolorbox}
\usepackage{longtable}
\newcommand{\mybox}[1]{
	\begin{tcolorbox}[
		boxsep=-2pt,
		standard jigsaw,
		boxrule=0.6pt,
		opacityback=0,
		sharp corners]
		#1
	\end{tcolorbox}
}

\newif\ifshowcomments
\showcommentsfalse

\ifshowcomments
\newcommand{\yao}[1]{\mytodored{[yao: #1]}}
\newcommand{\gu}[1]{\mytodocyan{[gu: #1]}}
\newcommand{\ye}[1]{\mytodoblue{[ye: #1]}}
\else
\newcommand{\yao}[1]{}
\newcommand{\gu}[1]{}
\newcommand{\ye}[1]{}
\fi

\newcounter{finding}
\newcommand{\finding}{%
  \stepcounter{finding}%
  \textbf{Finding} \textbf{\arabic{finding}}:%
}

\begin{document}
	
% \title{Boosting Pointer Analysis with Compiler-Based Offline Simplifications: An In-Depth Study}
\title{Accelerating C/C++ Pointer Analysis via Compiler-Based Offline Simplifications}

    %% Author with single affiliation.
\author{Zinan Gu}
%\authornote{with author1 note}          %% \authornote is optional;
                                        %% can be repeated if necessary
\orcid{xxxx}          %% \orcid is optional
\affiliation{
%  \position{Position1}
%  \department{Department1}              %% \department is recommended
  \institution{The State Key Laboratory of Blockchain and Data Security, Zhejiang University}            %% \institution is required
%  \streetaddress{Street1 Address1}
%  \city{City1}
%  \state{State1}
%  \postcode{Post-Code1}
  \country{China}                    %% \country is recommended
}
\email{xxxx@zju.edu.cn}        

\author{Peisen Yao}
\orcid{0000-0003-0342-9518}             %% \orcid is optional
\affiliation{
  \institution{The State Key Laboratory of Blockchain and Data Security, Zhejiang University}      
	\country{China}                   
}
\email{pyaoaa@zju.edu.cn}

\author{Kui Ren}
\orcid{0000-0003-3441-6277}             %% \orcid is optional
\affiliation{
  \institution{The State Key Laboratory of Blockchain and Data Security, Zhejiang University}       
	\country{China}                   
}
\email{kuiren@zju.edu.cn}

\begin{abstract}
Pointer analysis is a cornerstone of numerous static analysis applications, including compiler optimizations, slicing, bug detection, and verification.
While offline simplification is a common approach to boosting performance, existing methods are often tightly coupled to specific analysis algorithms and limited to a set of simplification rules.
This paper explores a new perspective: applying semantic-preserving compiler optimizations directly to intermediate representation (IR) before pointer analysis.
This strategy is modular, analysis-agnostic, and easily integrates with existing tools. We conduct an empirical study using diverse programs and three pointer analyses.
The results show substantial performance gains—up to $3.14\times$ speedup and $1.94\times$ memory reduction—while precision remains largely unchanged. 
We also analyze the trade-offs between optimization overhead and analysis speedup, quantify changes in IR structure, assess the characteristics of optimization configurations, and identify promising directions for future research.
\end{abstract}

% Our analysis reveals that optimization passes can be pruned by ~35\% on average while maintaining benefits, and identifies both universally beneficial and context-dependent passes.

%This study provides the first systematic investigation of compiler-based offline simplifications for pointer analysis, demonstrating their practical potential.

%Results show significant performance improvements—speedups up to 3.14× and memory reductions up to 1.94×—while precision remains largely unchanged.  %Custom optimization configurations often outperform standard O1-O3 levels.

\begin{CCSXML}
	<ccs2012>
	<concept>
	<concept_id>10003752.10010124.10010138.10010143</concept_id>
	<concept_desc>Theory of computation~Program analysis</concept_desc>
	<concept_significance>500</concept_significance>
	</concept>
	</ccs2012>
\end{CCSXML}

\ccsdesc[500]{Theory of computation~Program analysis}

\keywords{Pointer analysis, compiler optimizations, corpus study}
\maketitle

\section{Introduction}
\label{sec:introduction}

Pointer analysis is a class of static analysis techniques that determines the set of abstract locations a pointer may reference during program execution. The analysis results serve as a foundation for numerous static analysis applications, including compiler optimizations~\cite{lattner2007making,lattner2005automatic}, program slicing~\cite{li2016program,sridharan2007thin}, bug detection~\cite{sui2012static,shi2018pinpoint,tripp2009taj}, change-impact analysis~\cite{Orso:2004:ECD:998675.999453}, and formal verification~\cite{sea-dsa,wang2017partitioned,fink:typestate:issta}. 
Developing a practical pointer analysis requires addressing high-level conceptual issues, such as designing appropriate abstractions for the properties of interest, and handling low-level implementation details, such as selecting efficient data structures for the analysis.

The scalability of pointer analysis is critical for its practical application in large-scale, real-world programs.
Hence, there has been an everlasting interest in scaling pointer analysis, which can be roughly divided
into two classes: (1) maintaining precision while improving performance, such as sparsification~\cite{yan2018spatio}, pruning~\cite{fink:typestate:issta}, partitioning~\cite{kahlon2008bootstrapping}, parallelization~\cite{Mathew2010Parallel,Nagaraj2013Parallel, M2012AGPU,Nagaraj2013Parallel,zhao2018parallel,Edvinsson2011Parallel}, demand-driven analysis~\cite{heintze2001demand,zheng2008demand,sridharan2005demand,sridharan2006refinement,yan2011demand}), incremental analysis~\cite{liu2018d4,lu2013incremental},
 and database-based techniques~\cite{zuo2021chianina,shi2022indexing,Wang2017Graspan};
(2) trading precision and scalability, such as varying the level of analysis
sensitivities~\cite{kastrinis2013hybrid,smaragdakis2014introspective,jeon2018precise,li2018scalability,barbar2020flow} and using different flavors of heap abstractions (e.g.,  allocation-type abstraction~\cite{sundaresan2000practical} and recency abstraction~\cite{balakrishnan2006recency}). 
% for different program elements
% adaptive heap cloning~\cite{barbar2020flow},
%\yao{to improve}

In addition to algorithmic improvements, the efficiency of a pointer analysis is significantly influenced by the nature of the input provided to the analysis. 
To this end, various simplification techniques for pointer analysis are available, broadly classified into two categories: online and offline simplifications.  Online simplifications~\cite{Fahndrich1998,hardekopf2007ant,pearce2003online,lei2019fast} refer to transformations applied during the analysis, such as online cycle detection and collapsing. Offline simplifications~\cite{hardekopf2007exploiting,Rountev2000,Li2020,Lei2023}, on the other hand, preprocess the input data to reduce the problem size before the analysis. 
% Both approaches improve scalability while retaining precision.  
Despite the advances in simplification techniques, several limitations remain.  
\begin{itemize}
    \item First, existing simplifications are often tightly coupled to targeted pointer analysis algorithms, limiting their general applicability.
    %across different analyses. 
    For example, a simplification for flow-insensitive analysis~\cite{hardekopf2007ant,Rountev2000} may not be applicable to flow-sensitive or context-sensitive analyses. 
    \item Second, existing techniques operate over high-level abstract representations, such as symbolic pointer graph~\cite{yan2011demand},
    %  and pointer assignment graphs (PAG).
   %  inherently limiting the types of simplifications that can be performed. 
    The representations inherently limit the search space of simplification rules, such as merging equivalent vertices~\cite{hardekopf2007ant}, removing irrelevant edges~\cite{Li2020}, and filling graph nodes~\cite{Lei2023}.
    %, and rewriting grammars~\cite{TBD}.   
    \item Finally, current techniques perform the same strategies universally across all inputs,  resulting in missed opportunities for instance-specific simplifications.
  % do not fully account for variations in the input programs,
\end{itemize}
%\citet{Lei2023} introduce criteria for identifying foldable node pairs, leveraging an RSM-based graph-folding algorithm.

This paper introduces a new perspective on offline simplifications for pointer analysis.
Specifically, we leverage semantic-preserving transformations applied to programs' IR via compiler optimizations. The use of program transformations to improve static analysis has a long tradition, such as isolating hot paths~\cite{Ammons1998,Fisher1981,Melski2002}, rewriting loops~\cite{Balakrishnan2009,Gulwani2009,Sharma2011}, and simplifying arithmetic expressions~\cite{cooper2001operator}.
However, their application to pointer analysis remains relatively unexplored, particularly in the context of offline simplifications.
Specifically, for pointer analysis, compiler-based offline simplifications offer several advantages:

\begin{itemize}
    \item \emph{Broad Applicability}. Compiler optimizations can be applied broadly across pointer analyses with different precision (e.g., varying inflow- and context-sensitivities) and using different formulations (e.g., set constraints,  context-free language reachability, declarative languages, and type systems).
    \item \emph{Simplification Opportunities}. 
    Compiler optimizations operate directly on the IR, preserving the program's semantics and enabling more diverse and sophisticated simplifications. 
    %(e.g., inter-procedural analysis and transformations)
    For example, the \textsf{newgvn} optimization is flow-sensitive, handles both memory and scalar variables, and reasons about control and data flows, which goes beyond many existing graph simplification techniques.
    % which is absent in existing techniques. 
   % \yao{a bit unclear}
    \item \emph{Instance-specific Simplifications}. The effectiveness of simplifications can vary significantly depending on the characteristics of the input programs. 
    Compiler-based transformations can unlock additional performance gains that uniform approaches might miss by incorporating instance-aware simplifications.
\end{itemize}

From a practical standpoint, compiler-based offline simplifications are both modular and pluggable. First, they allow customization of simplifications for specific program components. For instance, there is growing interest in pre-analyzing libraries to enhance the scalability of whole-program analysis~\cite{rountev2008ide,arzt2016stubdroid,xiao2014persistent}. A library can be simplified once, and the results can be reused across various client programs and pointer analyses. Second, we can benefit from new advances in compiler optimizations without modifying the existing analysis algorithms.
Finally, compiler optimizations and conventional simplifications operate at different abstraction levels, and this separation opens opportunities for interaction between the two. For example, applying compiler optimizations could yield a simplified IR, which may enable more effective constraint graph simplifications.

While compilers hold promise as offline simplifiers for pointer analysis, their practical effectiveness remains unclear; the interactions between different optimizations are complex, and their combined effects are neither additive nor easily predictable. 
How do these transformations influence the performance of different pointer analyses? What are the trade-offs between the overhead introduced by these optimizations and their ability to improve scalability? What are some important optimization passes? 
How do they affect the precision of the analysis in exchange for scalability?
Understanding these trade-offs is crucial as they provide insights into the nature and potential of compiler-based offline simplifications.
%in the design space. Between optimization overhead and potential gains.

To address these questions, we evaluate compiler-based simplifications using $22$ open-source projects that vary in size and functionality. Specifically, we assess three state-of-the-art, inclusion-based pointer analyses for C/C++: DW-ander, a variant incorporating the wave propagation heuristic~\cite{pereira2009wave}; SCD-ander, a flow- and context-insensitive analysis with selective cycle detection; and VSFS~\cite{barbar2021object}, a flow-sensitive, context-insensitive analysis based on SFS~\cite{hardekopf2011flow} with object versioning~\cite{barbar2021object}. 
We systematically explore the compiler optimization space through iterative configuration search, evaluating both standard optimization levels (O1, O2, O3) and custom optimization sequences.
% across programs of varying complexity and domain characteristics.
 
Our results reveal several key findings about the end-to-end performance improvements  (\cref{subsec:end2end}), the effects on IR metrics (\cref{subsec:other-metrics}), the characteristics of optimization pass configurations  (\ref{subsec:passes}), and the impact of optimizations on analysis precision(\cref{subsec:precision}).
We also outline several directions for future work to extend the effectiveness (\cref{subsec:effectiveness}) and applicability (\cref{subsec:applicability}) of our approach.
Finally, during our evaluation, we identified twelve previously unknown bugs in SVF, several of which have since been addressed by the developers.
%mutating IR can also expose subtle bugs in analyzer implementations. 

To summarize, this paper makes the following key contributions:
\begin{itemize}
   \item We introduce a new perspective on offline simplifications for pointer analysis, leveraging semantic-preserving transformations applied to IR.
    \item We design an evaluation framework to systematically compare different simplification techniques. Our experiments on real-world programs and standard benchmarks provide valuable insights into the trade-offs between performance, precision, and overhead.
   \item We release the tools, implementation, and experimental data used in our study on our project homepage, enabling further research and replication of our results.
\end{itemize}

\section{Background and Motivation}
\label{sec:background}

This section discusses simplification in pointer analysis and motivates the use of compiler optimizations for pointer analysis.

\subsection{Graph Simplifications for Pointer Analysis}
\label{subsec:pta}
Static analysis plays a critical role in managing software complexity, but it is often hindered by the pervasive challenge of indirection, which complicates the analysis and diminishes its effectiveness. 
There are two predominant forms of indirection, namely indirect data flow (e.g., pointer dereferences in C, object field accesses in Java) and indirect control flow (e.g., function pointers in C, virtual method dispatch in Java, or closures in functional languages). 
%Even in languages that abstract away low-level pointers, such as those operating at the level of a compiler's IR, indirection is typically realized through pointer-based mechanisms.
Pointer analysis seeks to resolve indirection by approximating the memory locations a pointer may reference.  
Such information is crucial for the precision and performance of various downstream clients. 
For instance, pointer analysis facilitates various compiler optimizations, such as scalar promotion~\cite{lu1997register}, memory pool allocation~\cite{lattner2005automatic}, and auto-parallelization~\cite{vandierendonck2010paralax}. In software model checking~\cite{sea-dsa,wang2017partitioned}, pointer analysis constrains implicit dependencies between memory-stored values before more computationally expensive, path-sensitive analyses are applied. 
The efficiency and precision of pointer analysis directly impact these subsequent tasks. 

% Consequently, substantial research has been devoted to developing fast and precise algorithms.
% over the past several decades.
%\cite{shi2018pinpoint}

%fink:typestate:issta

%Pointer analysis aims to resolve this indirection by computing points-to sets. For each program entity (e.g., a variable or an object reference), its points-to set consists of the memory locations that may be indirectly accessed via that entity.

\begin{comment}

\begin{table}[t]
\centering
\caption{Examples of graph simplifications for pointer analysis.\yao{to fix}}
\label{table:simplifications}
\begin{tabular}{|l|l|l|}
\hline
\textbf{Type}   & \textbf{Analysis}    & \textbf{References}                                    \\ \hline
Online   & Flow-insensitive, context-insensitive   & \cite{Fahndrich1998,hardekopf2007ant,Rountev2000,hardekopf2007exploiting}                            \\ %\hline
& Flow-insensitive, context-sensitive      & \cite{}                                \\ \hline
Offline  & Flow-insensitive, context-insensitive & \cite{Rountev2000}                                     \\ %\hline
   & Flow-insensitive, context-insensitive                    & \cite{Li2020, Lei2023}        \\ \hline
\end{tabular}
\vspace{-3mm}
\end{table}
\end{comment}

Pointer analysis often uses graph-based representations to abstract various program constructs, such as the pointer expression graph (PEG)~\cite{zheng2008demand} and symbolic pointer graph (SPG)~\cite{yan2011demand}.
To improve its scalability, researchers have explored various graph simplification techniques: online simplifications~\cite{Fahndrich1998,hardekopf2007ant,pearce2003online,lei2019fast} dynamically adapt transformations during analysis based on the current state of the analysis, while offline simplifications~\cite{Rountev2000,hardekopf2007exploiting,Li2020,Lei2023} preprocess the input to reduce the problem size before the analysis begins.
Despite the progress, existing simplification techniques exhibit several limitations: First, they are mainly tailored to specific analyses, limiting their generalizability. Second, the abstract representations inherently limit the scope and effectiveness of simplifications. Third, they do not account for variations in input instances and apply uniform rules across all programs, missing instance-specific simplification opportunities.
%and use handcrafted rules

%The number of nodes collapsed is essential because it reduces both the number of nodes and the number of edges in the constraint graph; the more nodes that are collapsed, the smaller the input and the more efficient the algorithm
%These simplifications aim to reduce the size of graph representations by removing unnecessary variables or instructions, thereby accelerating analysis without sacrificing precision. 
%\yao{tbd: maybe add more papers for offline simp}

\subsection{Using Compiler Optimizations in Static Analysis}
\label{subsec:opt4spa}
%\yao{show more knowledge}

\begin{table}[t]
\caption{Examples of LLVM optimizations 
%\yao{to revise}
}
\label{tab:opt}
\centering
\begin{tabular}{| l  | l |}
%\toprule
\hline
\textbf{Granularity}  & \textbf{Typical optimization passes}    \\ \hline
Instruction   & \textsf{lower-switch}, \textsf{lower-invoke}, \textsf{lower-atomic}       \\ %\hline
%Instruction sequence  & \textsf{inst-combine}, \textsf{xxxx}, \textsf{xxxx}       \\ %\hline
Loop  & \textsf{loop-flatten}, \textsf{loop-interchange}, \textsf{loop-rotate}       \\ %\hline
Function  & \textsf{simplify-cfg}, \textsf{new-gvn}, \textsf{licm}       \\ %\hline
Module  & \textsf{ipsccp}, \textsf{lower-ifunc}, \textsf{wholeprogramdevirt}       \\ 
\hline
\end{tabular}
%\vspace{-3mm}
\end{table}

Compiler optimizations transform code to improve its efficiency without changing its semantics. These transformations can operate at various levels, from individual instructions to entire programs.
Table~\ref{tab:opt} outlines several standard optimizations in LLVM, categorized by their level of granularity. 
For example, transformations such as \textsf{lower-switch} and \textsf{lower-atomic} operate at the instruction level, while \textsf{simplify-cfg} and \textsf{newgvn} work on entire functions. 
Indeed, they have been widely used in static analysis for various goals, such as:
\begin{itemize}
    \item \emph{Normalization}. Compilers often standardize program representations, making it easier for static analyses.
    For example, the \textsf{loop-simplify} optimization restructures loops into a canonical form, facilitating loop analysis, and \textsf{merge-return} ensures that each function has a single return point, streamlining data flow analysis.
    Moreover, many analyses strictly depend on specific IR, such as Static Single Assignment (SSA)  or control-flow graphs (CFGs) without irreducible loops. For example, several algorithms for global value numbering~\cite{gargi2002sparse,rosen1988global} are explicitly designed to operate on SSA form. 
    %Indeed, they rely on some lightweight alias analysis in LLVM, such as \textsf{basicaa}, \textsf{tbaa}, and \textsf{globalsa}.

    \item \emph{Simplification}. 
    Compilers not only normalize but also simplify the code, making it more amenable to static analysis.
    There is substantial evidence that program transformations can significantly impact program analysis in both speed and precision.
    For example, algebraic simplifications such as constant folding and strength reduction can reduce the complexity of expressions, potentially reducing the analysis overhead~\cite{chen2018learning}. 
    Flow-insensitive pointer analysis may gain precision when applied to programs in SSA form, as SSA ensures each variable is assigned exactly once. 
\end{itemize}

%dead code elimination and global value numbering help reduce the number of variables and instructions that need to be tracked, improving analysis speed.
%For instance,  \textsf{mem2reg} converts memory-based accesses into register-based accesses, simplifying the analysis of variable lifetimes and dependencies. 

Another reason developers use compiler optimization to enhance static analysis is that it provides immediate access within standard development environments. 
For example, the Seahorn verification framework~\cite{Gurfinkel2015} leverages SMT-based model checking and abstract interpretation, both of which benefit from LLVM optimizations.
% Seahorn applies transformations, such as SSA conversion and dead-code elimination, during preprocessing to streamline verification. 
The Pinpoint static analysis framework~\cite{shi2018pinpoint} also applies a few optimizations, such as dead-code elimination, before running the path-sensitive bug-finder.
% function inlining, 

\subsection{Problem Statement}
\label{subsec:opt4pta}
%\yao{longer?}
The design space of pointer analysis approximations is vast and multi-dimensional, encompassing flow sensitivity, context sensitivity, field sensitivity, heap modeling, pointer representation, branch condition handling, and array indexing, among other factors. Identifying effective simplifications within this space remains a challenging problem.

This paper introduces a novel approach to offline simplification for pointer analysis by applying semantics-preserving compiler transformations directly to IR. Unlike traditional simplification techniques, the proposed method offers three key advantages. First, it is analysis-agnostic: the transformations apply uniformly across a wide range of pointer analyses, regardless of their underlying formulations or approximation strategies. Second, operating on the IR enables the approach to exploit rich semantic information, thereby opening new simplification opportunities. Third, the method supports program-specific optimizations, adapting transformations to the structural and behavioral characteristics of individual programs.

% Finally, they support modular simplifications, enabling pre-optimized libraries to be reused across multiple client programs and analyses.

Despite these potential benefits, several open questions remain regarding the practical effectiveness of compiler-based simplifications. How do these transformations affect different pointer analyses? What trade-offs exist between optimization overhead and scalability improvements? How do they impact analysis precision? In this work, we take a first step to address these questions through a systematic and comprehensive empirical evaluation.

\begin{comment}
   \begin{table}[t]
	\caption{Typical types of alias queries and their applications. \yao{if we do not evaluate these clients, do we need this table?}}
	\label{pointer-query}
	%\resizebox{0.7\textwidth}{!}
	{
		\begin{tabular}{  l l  l }
			\toprule
			\textbf{Query  Type}   & \textbf{Description}   & \textbf{Example}            \\ \midrule
			AliasPair($p$, $q$)  &If the pointer $p$ is an alias of $q$  & Race detection       \\
			PointsTo($p$)  & The points-to set for pointer $p$  &  Callgraph  \\ 
			PointedBy($o$) & The pointers that point-to memory $o$   & ??? \\ 
			AliasSet($p$)  &  The pointers that are aliased to pointer $p$ & Tainat analysis \\ \bottomrule
		\end{tabular}
	}
\end{table} 
\end{comment}

\section{Evaluation Methodology}
\label{sec:setup}

%\smallskip
\noindent \textbf{Research Questions}.
 In this paper, we formulate the following research questions to evaluate the impact of compiler-based offline simplifications for pointer analysis.
\begin{itemize} 
    \item \textbf{RQ1}: To what extent do compiler-based simplifications impact the performance of pointer analyses? 
    %Do they affect pointer analyses with different types similarly? 
    How do custom optimization configurations compare with standard compiler optimization levels?
    %\item \textbf{RQ2}: What underlying causes impact compiler-based simplifications on pointer analysis performance?\yao{This question can be dangerous} How do they modify the IR?
    %and constraint graphs?
    \item \textbf{RQ2}: What effects will compiler optimizations have on the characteristics of the IR, such as pointer counts and instruction counts?
    \item \textbf{RQ3}: What is the minimal set of passes required to achieve good performance gains? Which individual compiler optimization passes contribute most significantly to performance improvements? 
    % \yao{a new part}
   \item \textbf{RQ4}: Do the performance gains from compiler-based simplifications lead to significant variations in precision? 
   %\yao{Talking about precision can be dangerous: as the IR structures have changed significantly, it is challenging to determine whether the metrics are due to the IR or the algorithms.}}
\end{itemize}

% What is the trade-off between the overhead introduced by compiler-based simplifications and the performance gains in subsequent pointer analyses?  Which specific compiler optimizations have the most significant impact on the performance of pointer analysis?

%\begin{comment}
\begin{table}[t]
\caption{Definitions and terminology. An optimization configuration, denoted as $o$, is a set of compilation flags, each associated with a specific value.}
\label{tbl:name}
\centering
\begin{tabular}{| l | l | }
	\hline
			\textbf{Term} & \textbf{Definition} \\
			\hline
			Test programs & $p \in P$, where $|P| = 12$. \\
			Optimization configuration &  $o \in O$ where $|O| = 200$. \\
			Speedup of $o$ on program $p$ & $s_{o,p}$ \\
		Optimal speedup of program $p$ &  $s_{\text{optimal},p} = \max\{s_{o,p}, o \in O\}$. \\
		%Fraction of program optimal speedup of $o$ on $p$ &  $f_{o,p} = \frac{s_{o,p}}{s_{\text{optimal},p}}$. \\
		%	$o_{\text{opt}}$ with highest minimum fraction &   $o_{\text{opt}} = \arg\max_{o \in O} \min\{f_{o,p}, p \in P\}$. \\
		%	$o_{\text{opt}}$ with highest minimum speedup &  $o_{\text{opt}} = \arg\max_{o \in O} \min\{s_{o,p}, p \in P\}$. \\
		%	$o_{\text{opt}}$ with highest average fraction &  $o_{\text{opt}} = \arg\max_{o \in O} \text{mean}\{f_{o,p}, p \in P\}$. \\
		%	$o_{\text{opt}}$ with highest average speedup:  &  $o_{\text{opt}} = \arg\max_{o \in O} \text{mean}\{s_{o,p}, p \in P\}$. \\
\hline
\end{tabular}
\vspace{-3mm}
\end{table}
%\end{comment}
	
%\subsection{Pointer Analyzers}
\smallskip
\noindent \textbf{Pointer Analyzers}.
We have applied \ToolName\ (the pass sequence leading to the highest speedup) to evaluate three state-of-the-art and widely-used  pointer analyses for C/C++ varying in precision:
\begin{enumerate}
    \item DW-ander: Andersen analysis with the wave propagation heuristic~\cite{pereira2009wave}; 
    \item SCD-ander: Andersen analysis (inclusion-based, flow- and context-insensitive points-to analysis) with selective cycle detection;
%(2) VSFS~\cite{hardekopf2011flow}, an inclusion-based, flow-sensitive, context-insensitive points-to analysis;
   \item VSFS~\cite{barbar2021object}: inclusion-based, flow-sensitive, context-insensitive points-to analysis, based on SFS~\cite{hardekopf2011flow},  incorporating the object versioning technique in~\cite{barbar2021object};
\end{enumerate}

%(4) SeaDSA~\cite{gurfinkel2017context}, a unification-based, flow-insensitive, and context-sensitive points-to analysis with heap cloning, based on DSA~\cite{lattner2007making} and extended in~\cite{gurfinkel2017context}.
% For (1)-(3), 
We use the implementation maintained in  SVF~\cite{sui2016svf}\footnote{
\url{https://github.com/SVF-tools/SVF}}.
We focus on exhaustive pointer analysis and, thus, omit the demand-driven analysis SUPA~\cite{sui2016demand,sui2018vfdemand} in SVF.
%We cannot evaluate the pointer analyses in~\cite{guyer2005error,yu2010level,li2011boosting,li2013precise,sui2011spas,sui2014making,zhao2018parallel} because their implementations are not publicly available. 

%\footnote{\url{https://github.com/seahorn/sea-dsa}}.
%These analyses are field-sensitive, meaning each struct field is treated as a separate variable. 
%	\item SUPA~\cite{sui2016demand,sui2018vfdemand}, \footnote{\url{https://github.com/SVF-tools/SUPA}}	a demand-driven flow- and context-sensitive pointer analysis for C/C++. 
%	\item Canary, a unification-based, flow-insensitive, and context-insensitive points-to  analysis; 
%  %\yao{do we need a case study for FalconAA?...}

\begin{table}[t]
\caption{List of benchmarks and their characteristics. Lines of code (\textbf{LoC}) are obtained using cloc.}
\label{tbl:program}
\resizebox{0.9\textwidth}{!}
{
  \begin{tabular}{ | c | l | r | r | r | }
		%	\toprule
\hline
\textbf{Program} & \textbf{Description} & \textbf{LoC} & \textbf{Functions} & \textbf{Call Sites} \\ \hline
\texttt{ninja} & fast build system & 25.9K & 4.4K & 13.6K \\
\texttt{x264} & video encoding library & 72.9K & 0.9K & 4.8K \\
\texttt{tmux} & terminal multiplexer & 74.3K & 2.6K & 13.5K \\
\texttt{povray} & ray-tracing software & 81.2K & 2.1K & 17.7K \\
% kvrocks & distributed key-value store & 81.7K & 16.6K & 215.5K \\
\texttt{omnetpp} & network simulation  & 85.7K & 12.1K & 48.9K \\
\texttt{zsh} & customizable unix shell & 133.1K & 2.4K & 18.9K \\
\texttt{imagick} & image manipulation & 173.6K & 3.1K & 36.6K \\
\texttt{nginx} & HTTP web server & 176.5K & 1.8K & 10.7K \\
\texttt{sqlite} & relational database & 242.2K & 4.5K & 24.1K \\
\texttt{ctags} & source code indexer & 282.5K & 5.5K & 29.7K \\
\texttt{perl} & language interpreter & 292.4K & 3.1K & 27.1K \\
\texttt{xalancbmk} & XML transformation & 298.7K & 30.7K & 108.3K \\
\texttt{vvenc} & video encoding software & 348.7K & 17.8K & 97.6K \\
\texttt{thrift} & RPC framework & 350.2K & 6.6K & 145.0K \\
\texttt{bash} & shell program & 418.4K & 3.4K & 19.6K \\
\texttt{fish} & command shell & 482.3K & 22.7K & 71.3K \\
\texttt{trafficserver} & HTTP caching server & 693.0K & 32.4K & 120.4K \\
\texttt{gcc} & C compiler  & 978.3K & 26.8K & 180.8K \\
\texttt{vim} & text editing tool & $1,204.2$K & 7.0K & 49.2K \\
\texttt{boringssl} & SSL/TLS library & $1,681.2$K & 12.0K & 44.0K \\
\texttt{python} & language interpreter & $1,688.8$K & 13.4K & 82.3K \\
\texttt{php} & language interpreter & $2,848.6$K & 10.5K & 122.1K \\
\hline         
%\bottomrule
  \end{tabular}
}
%\vspace{-3mm}
\end{table}

\smallskip
\noindent \textbf{Benchmark Programs}.
Table~\ref{tbl:program} presents the benchmarks used in our experiments. We select $22$ open-source programs of varying sizes and functionalities. The programs range from a few thousand to nearly half a million lines of code and span diverse categories, including data compression, terminal management, HTTP server, code indexing, and RPC frameworks.
% \yao{sizes?}
%providing a comprehensive test suite for our evaluation.
%\yao{to revise}

\smallskip
\noindent \textbf{Compiler Optimizations}.
Our study employs the LLVM compilation framework
%the latest version compatible with both SVF and SeaDSA. \ye{tbd: Simplify this sentence, which is not connected so naturally:}
that organizes optimization passes at multiple levels of granularity, such as the module, function, and loop levels. It distinguishes between transformation passes, which include both default optimization pipelines and individual transformation passes
%(e.g., \textsf{module(concierge)}) \gu{seems not to exist} 
while excluding utility passes that do not perform optimizations (e.g., \textsf{dot-callgraph}) and transformations that are not semantics-preserving (e.g., \textsf{internalize}).
These optimization flags govern key and classic aspects of the compilation process, such as inlining, loop unrolling, and constant propagation. 
In total, we select \text{105} optimization flags to control these and other critical aspects of the compilation process. We rely on the pass manager to automatically inject dependent analysis passes as required.
%\yao{tbd} 
%Table xx lists all flags used. 

To explore the compiler optimization space, we employ a random optimization strategy. Specifically, we generate optimization sequences by randomly selecting flags, a well-established method in the compiler community. For each program, we evaluate $300$ distinct optimization sequences. 
%The sensitivity of our results to the number of sequences is analyzed in~\cref{sec:discuss}, where we extend the evaluation to up to \textbf{X} sequences. \yao{mei shi jian le}
%\yao{if we have enough time, maybe extend to more times for a few programs (following the PLDI 10 paper)}
Table~\ref{tbl:name} provides key definitions and notations used throughout this paper. While we introduce these definitions as needed in the text, the table serves as a convenient reference.

\smallskip
\noindent \textbf{Platform}.
All experiments are conducted on two servers, each equipped with
a 28-core Intel Xeon(R) Platinum 8176@2.10GHz  and $512$ GB of RAM, running Ubuntu 22.04.
We give each run a maximum of $12$ hours to complete and $128$ GB of memory.
% The results of timed-out and memory-out runs are still recorded.

%\subsection{Performance Measurements}
%\label{subsec:setup-measurement}
%We measure the time and overhead required to run the analyzer for each combination and program.  We explore the overall performance improvements, examine how these optimizations affect IR metrics, and identify the best combinations of compiler optimization flags that yield the most significant results.

%hardware performance counters through the xx interface, such as L1, L2, and TLB and branch prediction miss rates. We also collect 66 architecture-independent characteristics using the xx toolset, such as instruction mix, ILP, branch predictability, memory access patterns, and working set size.

\section{Compiler Optimizations Meet Pointer Analyses}
\label{sec:results-overall}
%This section examines the effects of compiler-based offline simplifications for pointer analysis. 
We first assess overall performance improvements using the configuration that disables all optimizations as the baseline (\cref{subsec:end2end}).
%Finally, we compare with different optimization levels (O1, O2, O3) of clang (\cref{subsec:passes}).
We then analyze the changes in IR metrics and analysis process (\cref{subsec:other-metrics}), 
explore the subtle impact of the passes (\cref{subsec:passes}),
and assess how the optimizations affect analysis precision (\cref{subsec:precision}).

%study the importance of the different optimization flags (\cref{subec:opt}),

%We then explore the causal relationships between optimizations and performance.
%using statistical methods to quantify their impact.

\subsection{End-to-End Improvements for the Analyses}
\label{subsec:end2end}

We evaluate the impact of compiler optimizations on the performance of pointer analyses. As a baseline, we compile programs with all optimizations disabled. We then measure the combined overhead of applying compiler optimizations and executing the pointer analysis.

\begin{figure}[htbp]
    \centering
    \includegraphics[width=0.9\textwidth]{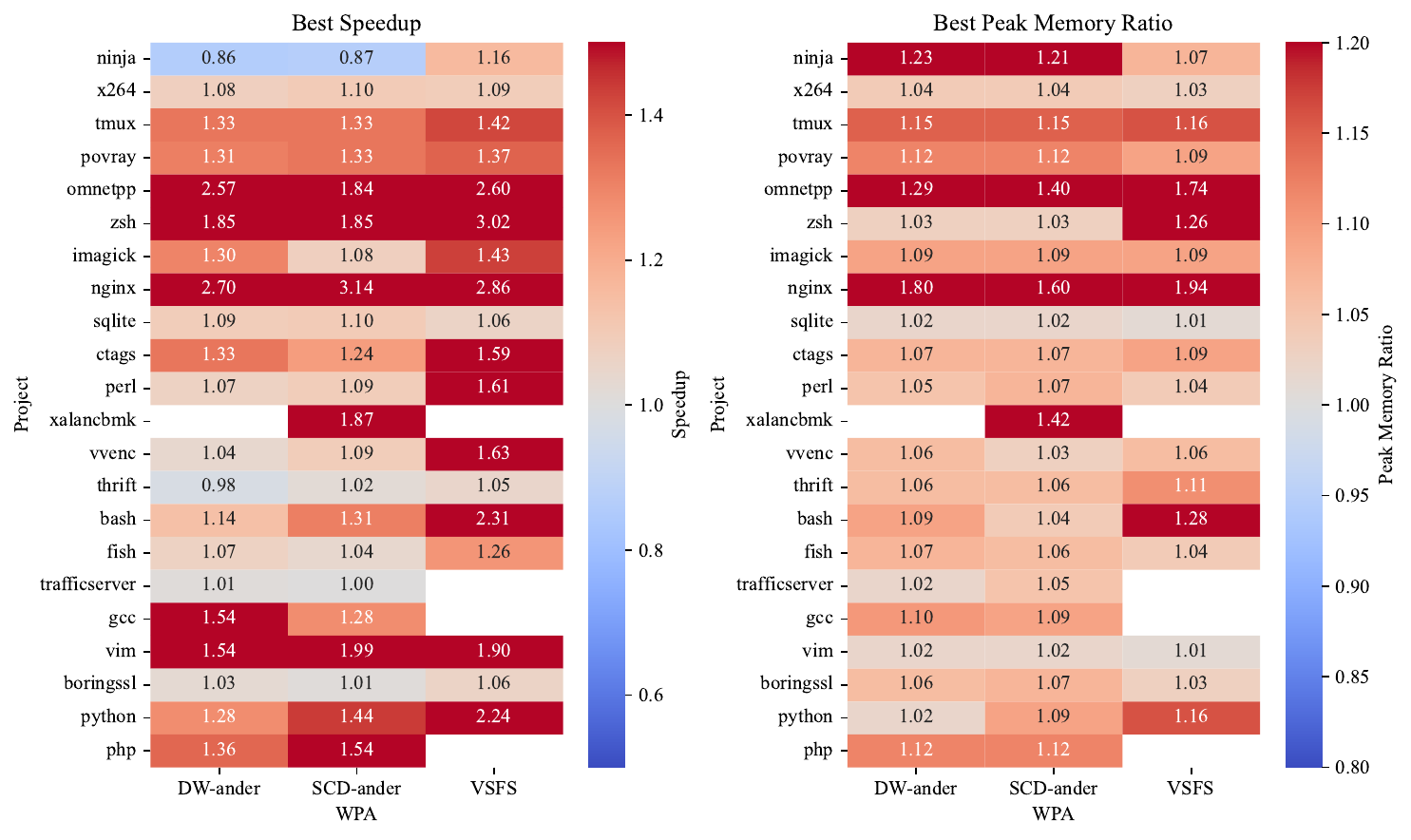}  % Replace with your PDF file
    \caption{The heat maps of best speedups and best peak memory ratios (thanks to \ToolName) for the three analyses and all the benchmarks.}
    \label{fig:heatmap-raw-double}
     \vspace{-3mm}
\end{figure}

\begin{figure}[htbp]
    \centering
    \includegraphics[width=0.9\textwidth]{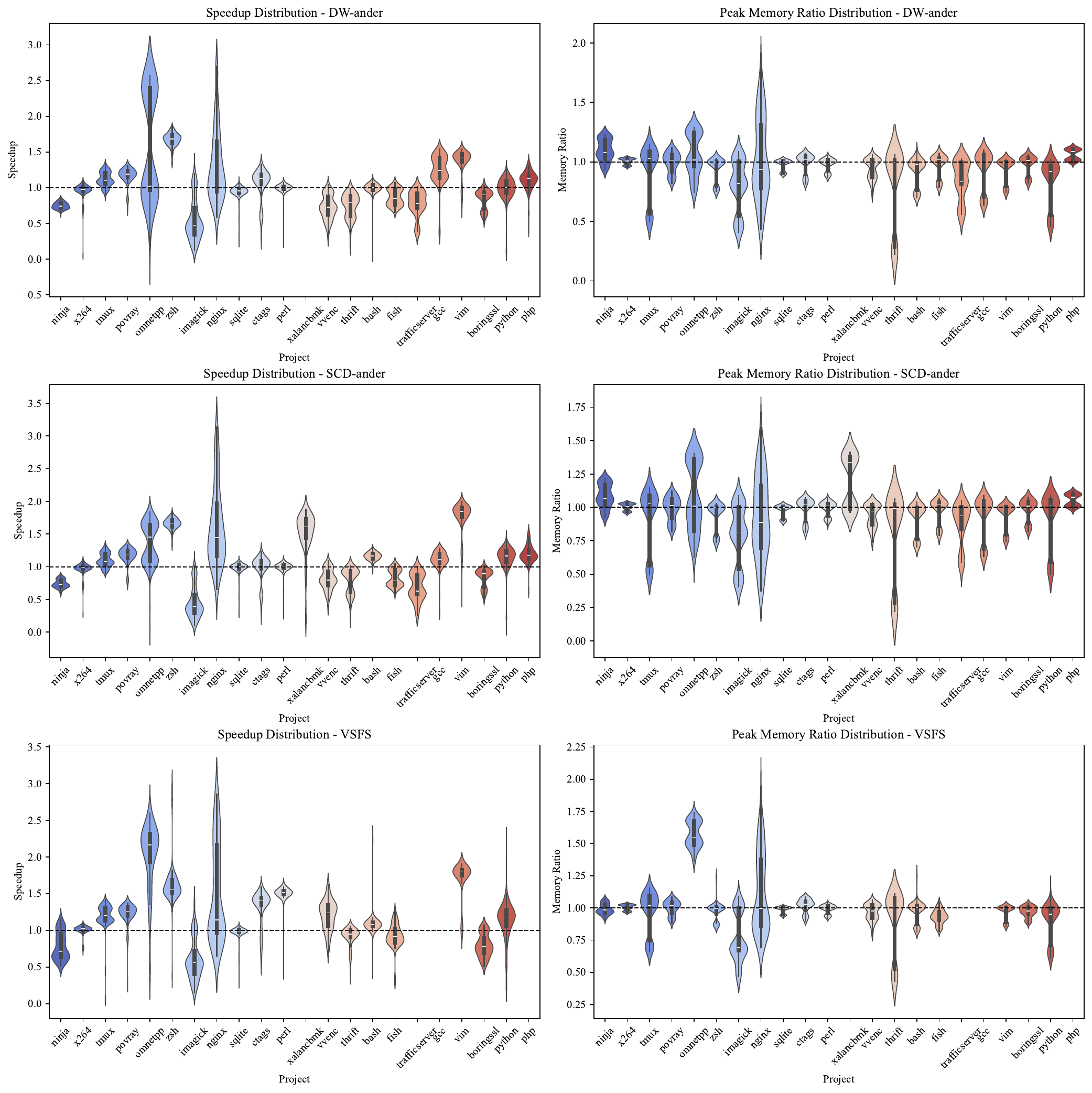}  % Replace with your PDF file
    \caption{Distribution of speedups and memory ratios across the benchmarks, where higher values are better.}
    \label{fig:violin-raw-double}
   % \vspace{-3mm}
\end{figure}

Table~\ref{tbl:speedups} reports execution times for both baseline and 
%optimized configurations
\ToolName, including the cost of optimization passes. Figure~\ref{fig:heatmap-raw-double} presents the maximum speedups, computed as the ratio of baseline to 
%optimized 
\ToolName\ execution time, along with the corresponding peak memory reductions. Figure~\ref{fig:violin-raw-double} shows the distribution of speedups and memory usage ratios across benchmarks.
% ordered by lines of code using violin plots.

\begin{table*}[t]
\centering
\caption{End-to-end improvements for analysis time (in \textbf{seconds}). 
DW-ander(\ToolName), SCD-ander(\ToolName), and VSFS(\ToolName) are the configurations with optimal speedups.}
\label{tbl:speedups}
\resizebox{0.9\textwidth}{!} 
{
\begin{threeparttable}
{
  \begin{tabular}{ | c | c | c | c | c | c | c | }
\hline
\textbf{Program}  & \textbf{DW-ander} & \begin{tabular}[c]{@{}c@{}}\textbf{DW-ander}\\ (\textbf{\ToolName})\end{tabular} & \textbf{SCD-ander} & \begin{tabular}[c]{@{}c@{}}\textbf{SCD-ander}\\ (\textbf{\ToolName})\end{tabular} & \textbf{VSFS} & \begin{tabular}[c]{@{}c@{}}\textbf{VSFS}\\ (\textbf{\ToolName})\end{tabular} \\  \hline
\texttt{ninja} & 9 & 10 & 7 & 9 & 40 & 34 \\
\texttt{x264} & 35 & 32 & 35 & 32 & 69 & 63 \\
\texttt{tmux} & 95 & 72 & 88 & 66 & 396 & 279 \\
\texttt{povray} & 58 & 44 & 52 & 39 & 268 & 196 \\
% kvrocks & \( 3,989 \) &  & \( 2,669 \) &  & OOT\tnote{1} & ?\tnote{2} \\
\texttt{omnetpp} & OOT\tnote{1} & \( 6,850 \) & \( 2,635 \) & 362 & OOT\tnote{1} & \( 15,958 \) \\
\texttt{zsh} & 198 & 107 & 182 & 98 & \( 1,648 \) & 546 \\
\texttt{imagick} & 49 & 38 & 37 & 34 & 179 & 125 \\
\texttt{nginx} & 384 & 142 & 462 & 147 & \( 2,768 \) & 969 \\
\texttt{sqlite} & 744 & 680 & 582 & 527 & \( 4,863 \) & \( 4,603 \) \\
\texttt{ctags} & 479 & 360 & 400 & 324 & \( 5,519 \) & \( 3,477 \) \\
\texttt{perl} & 814 & 758 & 700 & 640 & \( 5,636 \) & \( 3,502 \) \\
\texttt{xalancbmk} & OOT\tnote{1} & OOT\tnote{1} & \( 2,916 \) & \( 1,559 \) & OOT\tnote{1} & OOT\tnote{1} \\
\texttt{vvenc} & 750 & 722 & 756 & 696 & \( 16,426 \) & \( 10,106 \) \\
\texttt{thrift} & 141 & 144 & 114 & 112 & \( 1,204 \) & \( 1,145 \) \\
\texttt{bash} & 66 & 58 & 74 & 57 & 400 & 173 \\
\texttt{fish} & 132 & 124 & 120 & 115 & \( 1,586 \) & \( 1,258 \) \\
\texttt{trafficserver} & \( 2,724 \) & \( 2,703 \) & \( 1,903 \) & \( 1,906 \) & OOT\tnote{1} & OOT\tnote{1} \\
\texttt{gcc} & \( 4,471 \) & \( 2,905 \) & \( 3,540 \) & \( 2,765 \) & OOT\tnote{1} & OOT\tnote{1} \\
\texttt{vim} & \( 1,327 \) & 861 & \( 1,786 \) & 896 & \( 14,117 \) & \( 7,424 \) \\
\texttt{boringssl} & 318 & 307 & 313 & 311 & \( 4,335 \) & \( 4,073 \) \\
\texttt{python} & \( 1,625 \) & \( 1,272 \) & \( 1,862 \) & \( 1,290 \) & \( 16,243 \) & \( 7,241 \) \\
\texttt{php} & \( 4,975 \) & \( 3,660 \) & \( 5,794 \) & \( 3,755 \) & OOT\tnote{1} & OOT\tnote{1} \\

\hline
  \end{tabular}
  \begin{tablenotes}   
    \footnotesize              
    \item[1] OOT means the corresponding execution ran out of time, as a result.
    % \item[2] ? indicates experiments that were not completed or results that were not available, following the same. \gu{Maybe need to be modified.}
  \end{tablenotes}         
}
\end{threeparttable}
}
 % \vspace{-3mm}
\end{table*}

% \begin{figure*}
%     \centering
%     \includegraphics[width=0.85\textwidth]{images/heatmap-time-raw.pdf}  % Replace with your PDF file
%     \caption{The heat map of best speedups for the three analyses.}
%     \label{fig:heatmap-time-raw}
%      \vspace{-3mm}
% \end{figure*}

% \begin{figure*}
%     \centering
%     \includegraphics[width=0.75\textwidth]{images/heatmap-memory-raw.pdf}  % Replace with your PDF file
%     \caption{The heat map of best peak memory ratios for the three-pointer analyses.}
%     \label{fig:heatmap-memory-raw}
%     \vspace{-3mm}
% \end{figure*}

% \begin{figure*}[t]
%     \centering
%     \includegraphics[width=0.75\textwidth]{images/violin-time-raw.pdf}  % Replace with your PDF file
%     \caption{Distribution of normalized speedups across the programs}
%     \label{fig:violin-time-raw}
%     \vspace{-3mm}
% \end{figure*}

% \begin{figure*}[t]
%     \centering
%     \includegraphics[width=0.65\textwidth]{images/violin-memory-raw.pdf}  % Replace with your PDF file
%     \caption{Distribution of peak-memory ratios across the programs \yao{if the space is limited, maybe do not show memory?}}
%     \label{fig:violin-memory-raw}
%        \vspace{-3mm}
% \end{figure*}

\smallskip
\noindent \textbf{The Overall Results for the Analyses}. 
In summary, we observe the following key trends:
\begin{itemize}
    \item \emph{Impact on Time}.
    Compiler optimizations significantly reduced analysis time across most benchmarks. For instance, the execution time of \texttt{nginx} among all the $3$ kinds of analysis witnesses more than $60\%$ reduction, while that of \texttt{zsh} witnesses more than $45\%$. \ToolName\ reduces analysis time by up to about $35\%$ for \texttt{gcc} (DW-ander), while \ToolName\ achieves similar results for \texttt{ctags} (VSFS). However, not all optimizations result in consistent improvements—some configurations, such as those applied to Andersen analyses of \texttt{ninja}, introduced overheads that negate the potential performance gains. We should note that \texttt{ninja} has the fewest lines of code among all the programs, and its analysis is likely to benefit less from extensive optimization than larger programs.
    %Z%and suffer more from random disturbances. It took only $8$ seconds to finish the two Andersen analyses for \texttt{ninja}. Actually, \texttt{ninja} also did well in its VSFS(Opt) analysis.  
    Overall, the optimized configurations consistently outperform the baseline.
    % demonstrating the effectiveness of LLVM IR optimizations.

    \item \emph{Impact on Memory}.
    Memory consumption also decreases in some instances, though the reductions are generally less pronounced than those observed for execution time. For example, VSFS (\ToolName) analysis of \texttt{nginx} drastically reduces memory usage by about $50\%$, while SCD-ander (\ToolName) analysis of tmux demonstrates a moderate $13\%$ reduction. Despite these improvements, memory remains a limiting factor for most programs, where even optimized configurations struggle to scale gracefully.
  %  \ye{TODO}
    \item \emph{Differences among Benchmarks}. The impact of optimizations varies significantly across different programs. 
    While some benchmarks exhibit consistent speedups, others, such as \texttt{omnetpp} and \texttt{nginx}, show highly variable results, with certain optimizations improving performance by over $60\%$ while others lead to regressions. 
    % Interestingly, we observe no clear correlation between program size (measured in lines of code) and performance gains, suggesting that factors such as code complexity play a more significant role in determining the effectiveness of optimizations.
\end{itemize}

The performance improvements varied across different pointer analyses. Below, we break down the results for each analysis type.

\smallskip
\emph{DW-ander and SCD-ander}. 
    We first consider flow-insensitive, inclusion-based analyses that incorporate graph simplifications and constraint propagation into Andersen’s analysis.
    Compiler-driven offline simplifications further enhance performance. 
    DW-ander and SCD-ander show average speedups of $1.34$ and $1.38$, respectively, and up to $2.70$ and $3.14$ (both in \texttt{nginx}), respectively.
    For memory, compiler-based offline simplifications do not broadly reduce memory for DW-ander and SCD-ander, but they still work for a few Andersen analyses. 
    The improvements to SCD-ander are generally more significant than DW-ander, and SCD-ander is usually faster than DW-ander.
    For example, let us consider \texttt{omnetpp}.
    SCD-ander (\ToolName) is the fastest among the variants of the analyses, being $18.92$ $\times$ faster than DW-ander (\ToolName) and more than $44.08$ $\times$ faster than VSFS (\ToolName).
    %Without algorithmic optimizations, the original Andersen analysis (Ander) faced overheads that sometimes were even slower than the flow-sensitive analyses  (SFS, VSFS). In comparison, WAnder is xxx faster and xx than Ander.
    
\smallskip
\emph{VSFS}. We now consider the flow-sensitive and inclusion-based analysis.
    VSFS experiences the most significant reductions in memory consumption, with improvements of up to $1.94\times$. This is particularly relevant for memory-bound tasks, where reducing peak usage enables scaling to larger programs. Execution time also improves significantly, with VSFS achieving the highest speedups across the three analyses. Although VSFS exceeds time limits on a small subset of programs, this does not indicate a failure of compiler optimizations. Instead, it reflects the intrinsic computational cost of flow-sensitive analysis, which remains a fundamental bottleneck.
   Overall, VSFS benefits the most from optimizations in both time and memory, underscoring the effectiveness of compiler-driven transformations in mitigating the overhead of flow-sensitive pointer analysis.
  %  \yao{to extend}
    %\item \emph{SeaDSA}: The improvements in SeaDSA(Opt) were generally more moderate but consistent across all benchmarks. On average, SeaDSA(Opt) reduced analysis time by approximately xxx, with memory usage reductions in the same range. While the speedups were not as dramatic as those seen in Ander(Opt) or VSFS(Opt)\yao{to fix}

\mybox{\finding\ Compiler optimizations significantly reduce analysis time for most benchmarks, though results varied across them. Most benchmarks exhibit a mild reduction in memory usage, but some still show significant improvement. VSFS totally benefits most across the three pointer analyses.}

\begin{remark}
The SCD-ander incorporates online simplifications, such as selective cycle detection, while the DW-ander employs wave-propagation optimizations. Consequently, the performance improvements from compiler-based simplifications are additive.
\end{remark}

%(e.g., more than $100\%$ for \texttt{nginx})

\noindent \textbf{Balancing the Overhead and Benefits of Optimizations}.
% Figure~\ref{fig:balance} illustrates overhead introduced by optimization and the corresponding execution time. 
We find that the overhead introduced by optimization is negligible for nearly all programs, with the notable exception of \texttt{nginx}. In this case, the shortest execution time is $131.8$ seconds, while the optimization process takes up to $100$ seconds—resulting in a total time that significantly exceeds the best overall runtime observed.

\mybox{\finding\ Optimization overhead is negligible for most programs. We recommend using total time—including both optimization and execution—as the fitness function when searching for performant configurations to avoid a few exceptions.}

%\yao{maybe rm this part. (Optimization is usually fast/). Alternatively, we can briefly mention a few projects with high opt overhead.}
% To provide more insight into the trade-offs in the design space of compiler-based offline simplifications, we further study the performance benefits of optimizations and the costs of applying them. Given an analyzer $A$, an input program $p$, and a configuration of compiler optimizations $o$, we define the following metrics for measuring the effectiveness of optimizations
% \begin{itemize}
%     \item \text{Default analysis time $t$ and memory $m$}: The time and memory required for $A$ to analyze the original program $p$.
%     %without any optimizations.
%     \item \text{Optimization time $t_o$ and memory $m_o$}:  The time and memory taken by the compiler to apply the optimizations to $p$, producing a new, optimized input $p'$.
%     \item \text{Optimized analysis time $t'$ and memory $m'$}: The time and memory required for $A$ to analyze the optimized program $p'$.
% \end{itemize}
% Using these metrics, we define two key factors that reveal the impact of $o$ on the analysis: (1) Time overhead ratio: $t_o / (t - t')$;
% (2) Memory overhead ratio: $m_o / (m - m')$.

\begin{figure*}[t]
    \centering
    \includegraphics[width=0.9\textwidth]{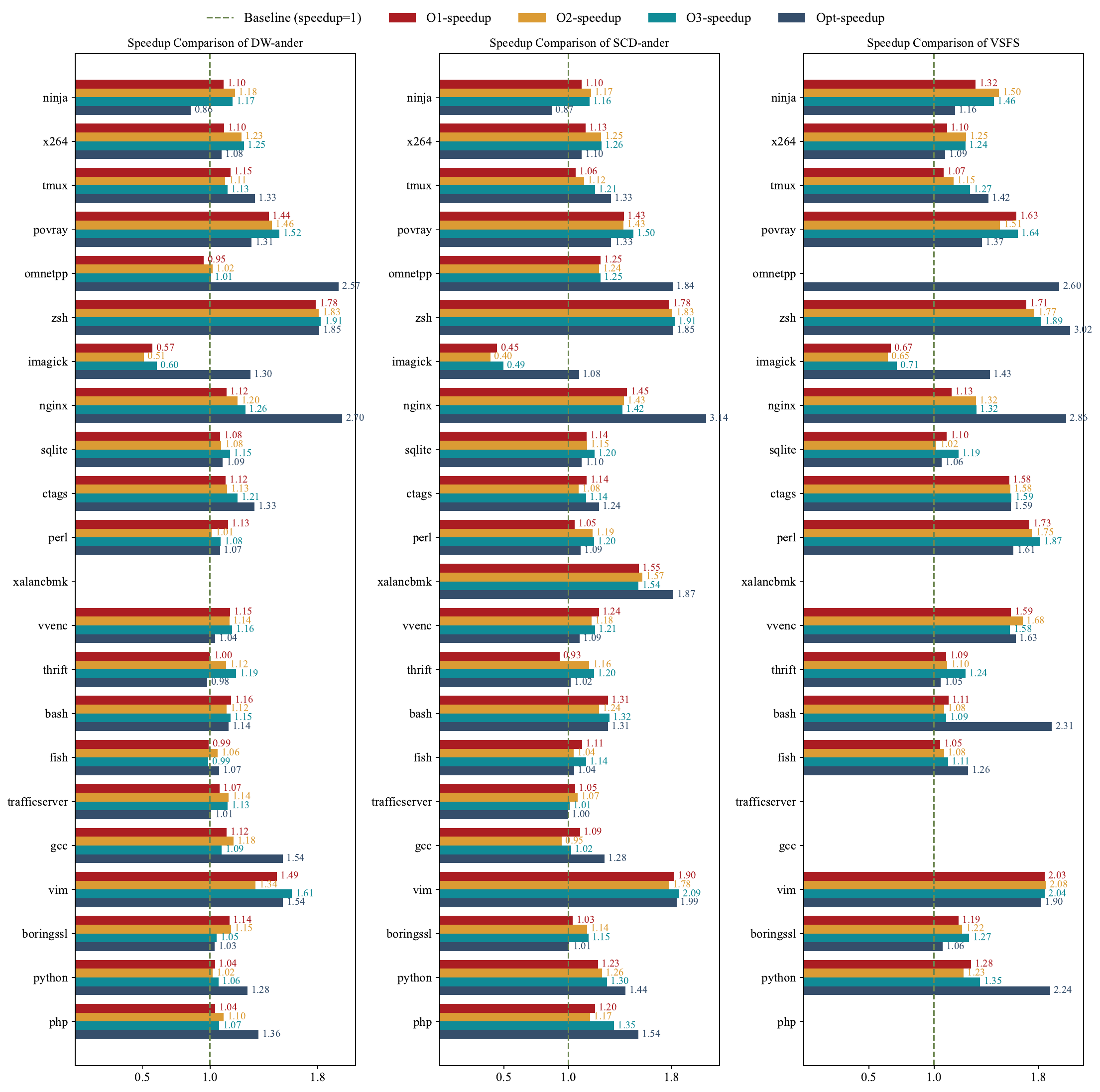}  % Replace with your PDF file
    \caption{Speedup comparison of O1, O2, O3, and \ToolName\ classified by analyses.}
    \label{fig:bar-vertical}
  %  \vspace{-3mm}
\end{figure*}

\smallskip
\noindent \textbf{Evaluating Standard Optimization Levels}. 
We now examine the impact of standard compiler optimization levels—O1, O2, and O3—on pointer analysis.  Figure~\ref{fig:bar-vertical} summarizes the results.
% These configurations are widely used in practice, yet their effects on analysis performance and precision remain understudied. 
% Specifically, we investigate whether they accelerate points-to analysis and how they influence precision. 

The standard optimization levels generally improve performance, with O3 often providing a marginal advantage over O1 and O2. Notably, \texttt{vim} (VSFS) exhibits consistent speedups across all optimization levels ($2.032$ under O1, $2.081$ under O2, and $2.040$ under O3). Conversely, some programs, such as \texttt{imagick}, experience slowdowns (e.g., VSFS peaks at $0.71$ under O3), indicating that standard optimization levels do not universally benefit pointer analysis.

Comparing these results to our iterative optimization approach, we observe that \ToolName\ outperforms O1, O2, and O3 in most cases. However, in certain instances, standard compiler optimizations yielded better results. These findings highlight the potential of compiler optimizations in accelerating pointer analysis and underscore the need for optimization strategies tailored to individual programs. Further research into selecting appropriate optimization passes could enhance performance beyond standard configurations. %\gu{textsf?}

 %suggesting that general-purpose optimizations can be effective but may not be optimal for all programs
 
% \begin{figure*}[t]
%     \centering
%     \includegraphics[width=0.65\textwidth]{images/bar-horizontal.pdf}  % Replace with your PDF file
%     \caption{Speedups classified by optimizations.}
%     \label{fig:bar-horizontal}
%     \vspace{-3mm}
% \end{figure*}

%Figure~\ref{fig:bar-horizontal} shows speedups organized by Optimizations, with which we can clearly compare the speedups of three points-to analysis techniques under the same optimization. In total, the acceleration of VSFS ranked first among the three techniques and was better than each of the optimizations. Although VSFS might lag sometimes, the gap was always narrow. This indicates that VSFS benefited the most from compiler optimizations. DW-ander and SCD-ander also performed well in many cases and often showed few differences in speedups, but there were exceptions like \texttt{omnetpp} (Opt, DW was much better SCD) and \texttt{vim} (all, SCD was much better than DW). 

% Deeper research is required to determine factors contributing to these phenomena, and it may be helpful to choose proper passes when trying to find better combinations.

\mybox{\finding\ Although standard optimizations accelerated pointer analysis in some cases, many still benefited significantly more from \ToolName. Relying on standard optimization levels may inadvertently obscure superior alternatives.}
\subsection{Effects on the IR  Metrics}
\label{subsec:other-metrics} 

\begin{figure}
    \centering
    \includegraphics[width=0.9\textwidth]{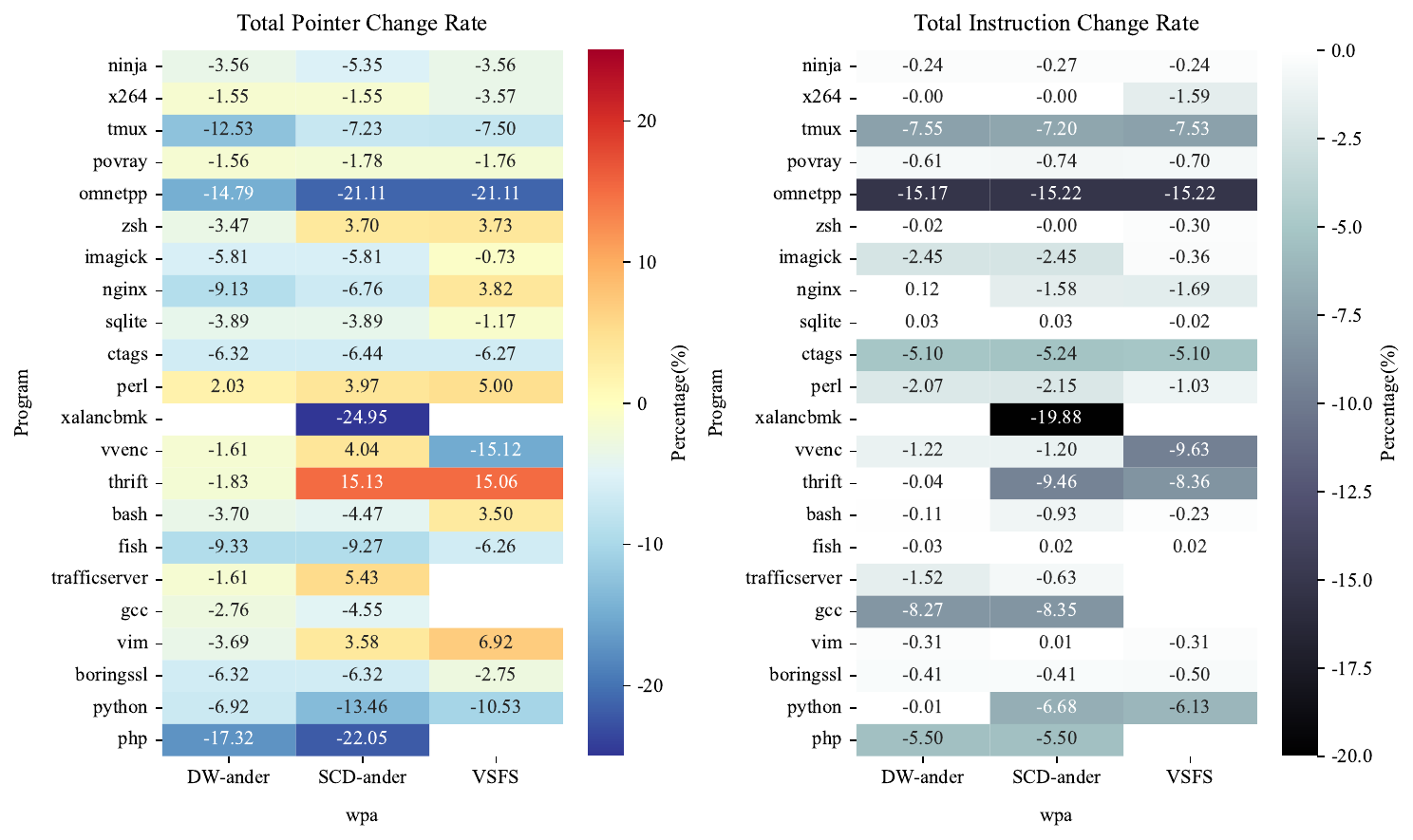}  % Replace with your PDF file
    \caption{Heat maps of total pointer change rates and total instruction change rates.}
    \label{fig:heatmap-metric-change-rate}
  %  \vspace{-3mm}
\end{figure}

To better understand how compiler-driven offline simplifications influence pointer analysis performance, we examine their effects on IR metrics. 
% Compiler optimizations can significantly alter the IR structure, affecting performance in pointer analyses. 
Specifically, we examine whether these optimizations affect the elimination of variables and instructions. 
Figure~\ref{fig:heatmap-metric-change-rate} summarizes the results.
% constraint graph sizes, and the profiling data of pointer analyses.

\smallskip
\noindent \textbf{Pointers Eliminated}. 
One significant benefit of compiler optimizations is the elimination of redundant or unnecessary pointers. For most benchmarks, the number of pointers after optimization is lower than the corresponding baseline (according to the left part of Figure~\ref{fig:heatmap-metric-change-rate}). For example, compiler optimizations reduce pointer counts from $431,145$ to $340,115$ for the VSFS analysis of \texttt{omnetpp} benchmark, by approximately $21.11\%$. This reduction is especially beneficial for flow-sensitive pointer analyses, such as VSFS, where facts must be tracked by distinguishing program points. In some cases (e.g., \texttt{perl} and \texttt{vim}), the optimized counts exceed the baselines, but such a phenomenon is uncommon.

\smallskip
\noindent \textbf{Instructions Eliminated}.  
Compiler optimizations also affect the number of instructions in IR.
According to the right part of figure~\ref{fig:heatmap-metric-change-rate}, instruction counts after optimization are slightly lower than baselines in most cases, but there are still a few programs, the instruction counts of which drop significantly. For instance, the benchmark \texttt{xalancbmk} experiences a notable reduction from $796,964$ to $638,515$ during optimization for SCD-ander. Although instruction counts do not drop so drastically as pointer counts, none rose significantly. 
%In some cases, compiler optimizations can reduce instruction counts.

\begin{figure}[t]
    \centering
    \includegraphics[width=0.9\textwidth]{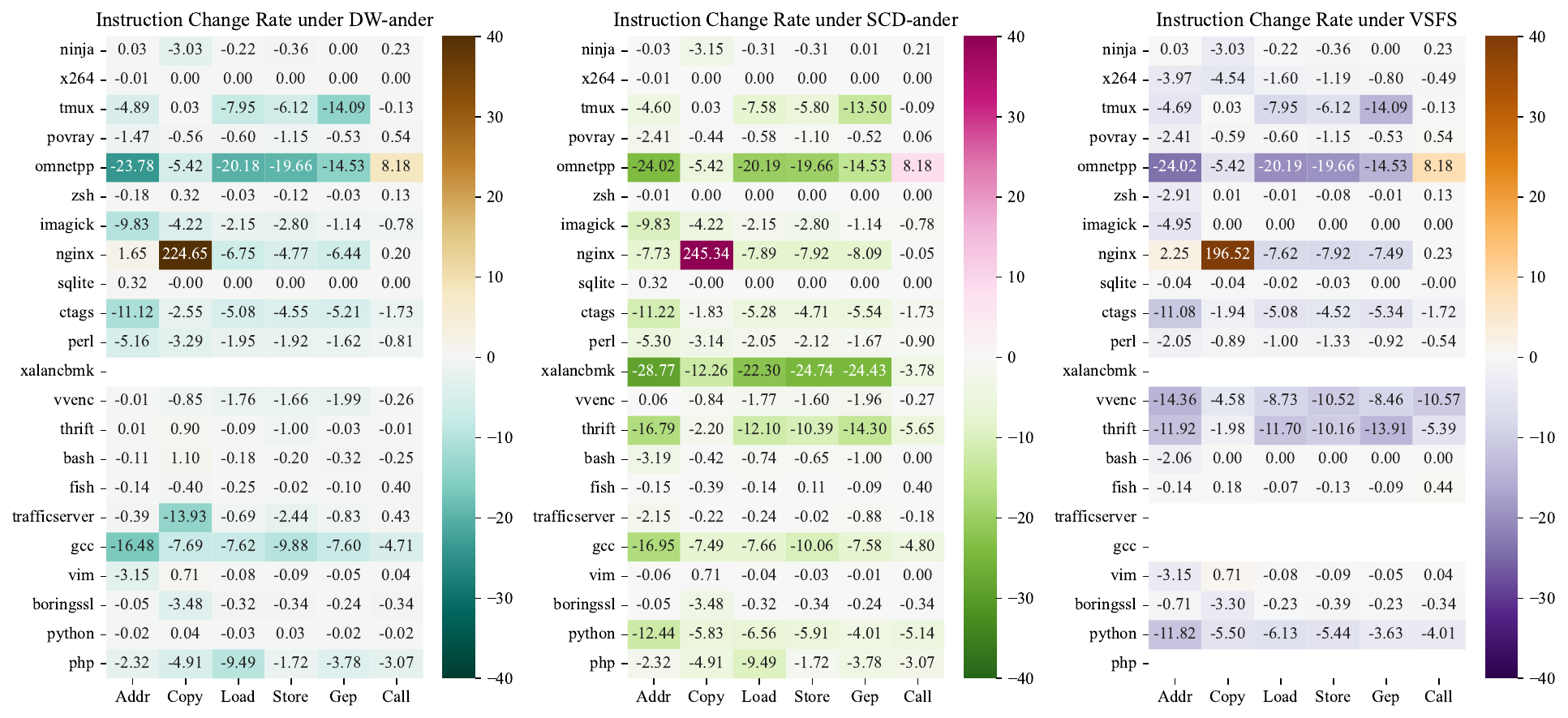}  % Replace with your PDF file
    \caption{The heat map of separate instruction ratios.}
    \label{fig:heatmap-instruction-change-rate}
    %\vspace{-3mm}
\end{figure}

\smallskip
\noindent \textbf{Separate Instruction Variation}. 
%\yao{no numbers mentioned}
Figure~\ref{fig:heatmap-instruction-change-rate} shows separate instruction change rates between baselines and {\ToolName}, including Addr, Copy, Load, Store, Gep, and Call. The ratios usually fell slightly, but there were also significant decreases in specific programs such as \texttt{omnetpp} and \texttt{xalancbmk}. 
% The total instruction ratios of these programs also became more prominent and significant than most benchmarks.
\texttt{Nginx} is an interesting case, where Copy rose significantly by $224.65\%$ (DW-ander), $245.34\%$ (SCD-ander), and $196.52\%$ (VSFS), while Load and Store dropped slightly; however, the total counts did not appear to vary as drastically as the separate counts, which means Load and Store might be replaced by Copy. Referring to the original data, we find that the Copy instructions of \texttt{nginx} are significantly smaller than those of the other instructions before optimization. 
% The subtle variation in Addr, Load, Store, GEP, and Call offsets the fierce rise of Copy. 
%This indicates that compiler optimization might transform Load, Store, Gep, or Call into Copy for \texttt{nginx}. 
Moreover, in time and memory consumption, \texttt{nginx} stands out. It is worth studying whether a special relationship exists between such transformation and points-to analysis overhead in the future. 
We find something interesting after further experiments, and we will detail relevant findings in \cref{subsec:passes}. 
 % \yao{why not study here? need to study in the future?}

%Overall, neither pointer counts nor instruction counts fluctuated significantly. Total pointer ratios range from $0.9$ to $1.3$, and total instruction ratios range from $1.0$ to $1.2$.

% Interestingly, benchmarks like \texttt{zsh} and \texttt{vim} saw increased pointer counts yet still achieved substantial speedups, indicating that structural simplifications beyond raw counts also play a key role in performance.

%\ToolName\ reduces both pointer and instruction counts across most benchmarks, typically resulting in faster pointer analysis. 

\mybox{\finding\ While pointer and instruction counts offer a coarse approximation of IR complexity, they do not fully account for the factors that influence analysis cost. Performance is also affected by the size of points-to sets, the dynamics of the worklist algorithm, and the structure of value-flow graphs in flow-sensitive analyses.}

\subsection{Analysis of Optimization Passes}
\label{subsec:passes}

%\yao{we should analyze more (pruning, pass importance, clustering, causality...)}
In this section, we take a closer look at the structure and effectiveness of the optimization passes. The optimization configurations generated by  \ToolName\ typically contain over $50$ compiler flags, making it difficult to determine which passes are essential for performance. To address this, we conduct two complementary experiments: (1) pruning redundant passes from optimized configurations, and (2) evaluating the isolated impact of individual passes. We exclude \texttt{ninja} and \texttt{x264} from the experiments due to their insufficient scale.

\smallskip
\noindent \textbf{Pruning Redundant Passes}. 
We begin by pruning each optimized configuration produced by \ToolName. Passes are removed one at a time in random order. A pass is discarded if its removal does not increase analysis time by more than $1\%$; the baseline is updated after each successful removal. This process continues until no further passes can be eliminated without exceeding the threshold.

%\yao{``size'' of the final configurations? e.g., how many options they contain before and after the reduction...}

\begin{figure}[t]
    \centering
    \begin{minipage}[t]{0.49\textwidth}
        \centering
        \includegraphics[width=\textwidth,height=7cm,keepaspectratio]{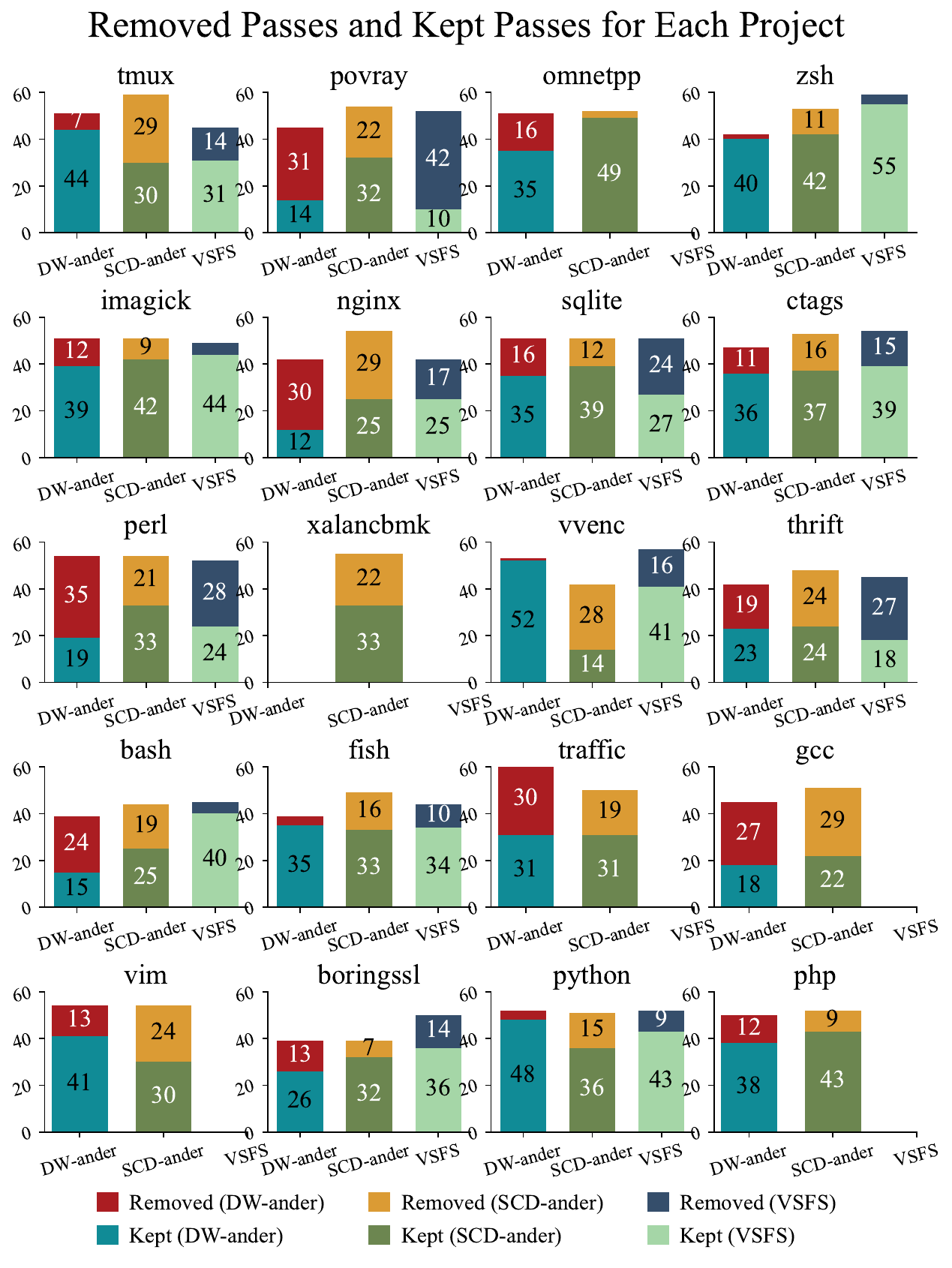}
        \caption{Removed and kept passes.}
        \label{fig:heatmap-opt-pruned}
    \end{minipage}
    %\hfill
    \begin{minipage}[t]{0.48\textwidth}
        \centering
        \includegraphics[width=\textwidth]{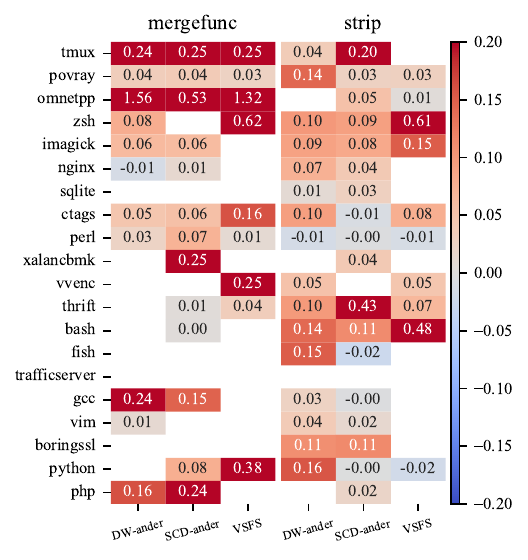}
        \caption{Time variation with a certain optimization pass disabled.}
        \label{fig:beneficial-heatmaps}
    \end{minipage}
   % \vspace{-3mm}
\end{figure}

\begin{comment}
    \begin{figure}[t]
    \centering
    \includegraphics[width=0.5\textwidth]{images/opt-pruned.pdf}  % Replace with your PDF file
    \caption{Removed Passes and Kept Passes for Each \ToolName.}
    \label{fig:heatmap-opt-pruned}
    \vspace{-3mm}
\end{figure}

\begin{figure}
    \centering
    \includegraphics[width=0.5\textwidth]{images/beneficial_heatmaps_2.pdf}  % Replace with your PDF file
    \caption{Heat maps of some beneficial passes.}
    \label{fig:beneficial-heatmaps}
    \vspace{-3mm}
\end{figure}
\end{comment}

Figure~\ref{fig:heatmap-opt-pruned} shows the number of passes removed and retained across benchmarks. On average, pruning reduces the size of optimization sequences by $34.6\%$, with some configurations shrinking by over $50\%$. Every one of the $105$ available passes is removed in at least two configurations. These results indicate that \ToolName\ frequently produces over-specified and sometimes even contains detrimental passes. \gu{add detrimental} Accordingly, a significantly smaller subset of passes suffices to achieve near-optimal performance.
%\yao{first point: to what extent the sizes are reduced}
%\gu{to count}
%\yao{any second point?}
%For example, the \texttt{-fast} optimization level includes inlining and vectorization by default, but both optimizations can sometimes degrade performance. The iterative optimization approach more effectively applies these optimizations to specific programs, for instance, applying inlining selectively to programs with small functions or disabling vectorization when it proves detrimental to performance.\yao{seems to have copied from another paper? (and not very relevant?)}

\smallskip
\emph{Frequently Beneficial Passes}.
Some passes consistently improve performance across programs and analyses. Two notable examples are shown in Figure~\ref{fig:beneficial-heatmaps}. If they are pruned, the time overhead often increases.
For instance, disabling \textsf{mergefunc} frequently causes significant slowdowns, particularly in large codebases. This pass merges semantically equivalent functions, even when they differ syntactically at the IR level, thereby reducing function count and simplifying alias analysis.
%\yao{add data}
Similarly, the \textsf{strip} pass removes symbolic metadata, such as debug information and internal symbols. This simplification reduces the number of entities and constraints tracked during analysis, yielding consistent speedups. %, especially in smaller benchmarks.
%\yao{why does debug info affect ...? a bit strange}

%\scriptsize
\begin{table}[t]
\centering
\caption{Part of context-dependent passes. Percentages represent the change rates over time after corresponding passes are removed.}
\label{tab:context-dependent passes}
\resizebox{0.9\textwidth}{!}
{
\begin{tabular}{|c|lcr|lcr|}
\hline
\textbf{Pass} & \multicolumn{3}{|c|}{\textbf{Positive}} & \multicolumn{3}{|c|}{\textbf{Negative}} \\
\hline
\multirow{2}{*}{\textsf{sroa}} & \texttt{fish} & VSFS & $+8.07\%$ & \multirow{2}{*}{\texttt{bash}} & \multirow{2}{*}{VSFS} & \multirow{2}{*}{$-5.13\%$} \\
& \texttt{nginx} & SCD-ander & $+21.16\%$ & & & \\
\hline
\textsf{slsr} & \texttt{python} & VSFS & $+10.77\%$ & \texttt{bash} & VSFS & $-8.73\%$ \\
\hline
\multirow{3}{*}{\textsf{scalarizer}} & \texttt{php} & SCD-ander & $+11.87\%$ & \multirow{3}{*}{\texttt{bash}} & \multirow{3}{*}{VSFS} & \multirow{3}{*}{$-21.42\%$} \\
& \texttt{python} & DW-ander & $+10.58\%$ & & & \\
& \texttt{python} & VSFS & $+8.64\%$ & & & \\
\hline
\multirow{2}{*}{\textsf{globalsplit}} & \texttt{php} & SCD-ander & $+11.97\%$ & \multirow{2}{*}{\texttt{boringssl}} & \multirow{2}{*}{SCD-ander} & \multirow{2}{*}{$-5.90\%$} \\
& \texttt{python} & DW-ander & $+14.51\%$ & & & \\
\hline
\textsf{nary-reassociate} & \texttt{imagick} & VSFS & $+10.27\%$ & \texttt{fish} & VSFS & $-9.59\%$ \\
\hline
\multirow{3}{*}{\textsf{gvn}}& \multirow{3}{*}{\texttt{traffic}} & \multirow{3}{*}{DW-ander} & \multirow{3}{*}{$+16.00\%$} & \texttt{nginx} & DW-ander & $-16.47\%$ \\
& & & & \texttt{nginx} & SCD-ander & $-20.00\%$ \\
& & & & \texttt{nginx} & VSFS & $-20.41\%$\\
\hline
\textsf{lower-} & & & & & & \\
\textsf{widenable-} & \texttt{php} & DW-ander & $+10.47\%$ & \texttt{python} & SCD-ander & $-5.28\%$ \\
\textsf{condition} & & & & & & \\
\hline
\multirow{3}{*}{\textsf{barrier}} & \texttt{fish} & DW-ander & $+8.07\%$ & \multirow{3}{*}{\texttt{tmux}} & \multirow{3}{*}{VSFS} & \multirow{3}{*}{$-7.23\%$} \\
& \texttt{imagick} & VSFS & $+12.80\%$ & & & \\
& \texttt{python} & DW-ander & $+13.32\%$ & & & \\
\hline
\multirow{3}{*}{\textsf{attributor}} & \texttt{boringssl} & VSFS & $+11.57\%$ & \multirow{3}{*}{\texttt{vvenc}} & \multirow{3}{*}{VSFS} & \multirow{3}{*}{$-10.83\%$} \\
& \texttt{python} & DW-ander & $+10.73\%$ & & & \\
& \texttt{traffic} & DW-ander & $+14.36\%$ & & & \\
\hline
\textsf{inline} & \multicolumn{3}{|c|}{$\ge 7$} & \texttt{zsh} & VSFS & $-15.07\%$ \\
\hline
\end{tabular}
}
%\vspace{-3mm}
\end{table}

\normalsize
\smallskip
\emph{Context-Dependent Passes}.
Several optimization passes exhibit highly variable effects on performance, contingent on both the program and the analysis. Table~\ref{tab:context-dependent passes} reports passes whose removal leads to both improvements and regressions in analysis time. Each entry shows the relative change in analysis time when a pass is disabled: positive values indicate slowdowns, while negative values indicate speedups.
Notably, many of these passes degrade the performance of the VSFS analysis, suggesting that VSFS is particularly sensitive to compiler optimization changes. These observations highlight the need for adaptive optimization strategies that consider both the characteristics of the target program and the sensitivity of the downstream analysis.

\mybox{\finding\ 
Many passes in the optimization configurations can be pruned without degrading performance. While some passes provide consistent benefits, others exhibit context-dependent behavior. Flow-sensitive analyses, such as VSFS, are particularly susceptible to performance degradation from suboptimal transformations.
%Optimization sequences produced by \ToolName\ are often over-specified: many passes can be pruned without degrading performance. While some passes provide consistent benefits, others exhibit context-dependent behavior that varies with program structure and analysis sensitivity. Flow-sensitive analyses, such as VSFS, are particularly susceptible to performance degradation from suboptimal transformations.
}

\smallskip
\noindent \textbf{Exploring the Individual Passes}.
While pruning identifies passes that are necessary in combination, it does not attribute performance gains to individual passes due to interactions among them. To isolate effects, we apply each pass independently to unoptimized bitcode and measure its impact on analysis time and IR metrics. Due to scalability constraints, this experiment is performed on a representative subset of benchmarks, such as \texttt{nginx}.

%\smallskip
%\emph{Case Study: Load-Store Vectorizer}. 
%\yao{Maybe this part can be a bit strange. Is it counterintuitive? Does it bring some insight?}
%Several findings emerge. Notably, 
\texttt{Nginx} exhibits a strong dependency on the \textsf{load-store-vectorizer} pass. When this pass is removed from the optimized configuration, analysis time increases by $135.06\%$, $105.10\%$, and $213.85\%$ for DW-ander, SCD-ander, and VSFS, respectively. Simultaneously, the number of Copy instructions in the IR decreases by $68.70\%$, $56.27\%$, and $67.81\%$. Conversely, when the pass is applied in isolation, analysis time decreases by $57.58\%$, $64.08\%$, and $63.29\%$, while the number of Copy instructions increases by $227.07\%$. These results indicate that \textsf{load-store-vectorizer} is both critical for performance and the primary source of Copy instructions in \texttt{nginx}.

This behavior stems from the vectorizer's type-correction logic. When vectorizing scalar loads of heterogeneous types (e.g., \verb|float| and \verb|i32*|), the pass coerces them into a uniform vector type (e.g., \verb|i32|). Upon extraction, scalar values inherit the uniform type, necessitating bitcasts to restore original kinds. Since SVF models bitcasts as Copy operations, this transformation inflates the number of Copy instructions.
We observe similar patterns in other benchmarks. For example, for \texttt{vvenc} (VSFS), removing the \textsf{load-store-vectorizer} increases analysis time by $19.51\%$, while Copy instructions decrease by only $0.74\%$. In the experiments of individual passes, Copy instructions increase by $1.26\%$ and VSFS time decreases by $4.19\%$ with this pass working alone. Although the effect is less pronounced than that in \texttt{nginx}, no other pass removal produces a comparable trend, suggesting a consistent underlying mechanism. \gu{The single experiment of vvenc (VSFS) is running.}

\mybox{\finding\ Optimization passes can significantly accelerate analysis, yet they concurrently introduce intricate variations to IR. The subtlety of their underlying mechanisms can often yield counterintuitive outcomes.
%Optimization passes are a double-edged sword: they dramatically speed up analysis but also add sophisticated layers of complexity to the IR. Their subtle internal workings can trigger counterintuitive effects.
}

\subsection{The Impact on the Analysis Precision}
\label{subsec:precision}
\begin{figure}[t]
    \centering
    \includegraphics[width=0.85\textwidth]{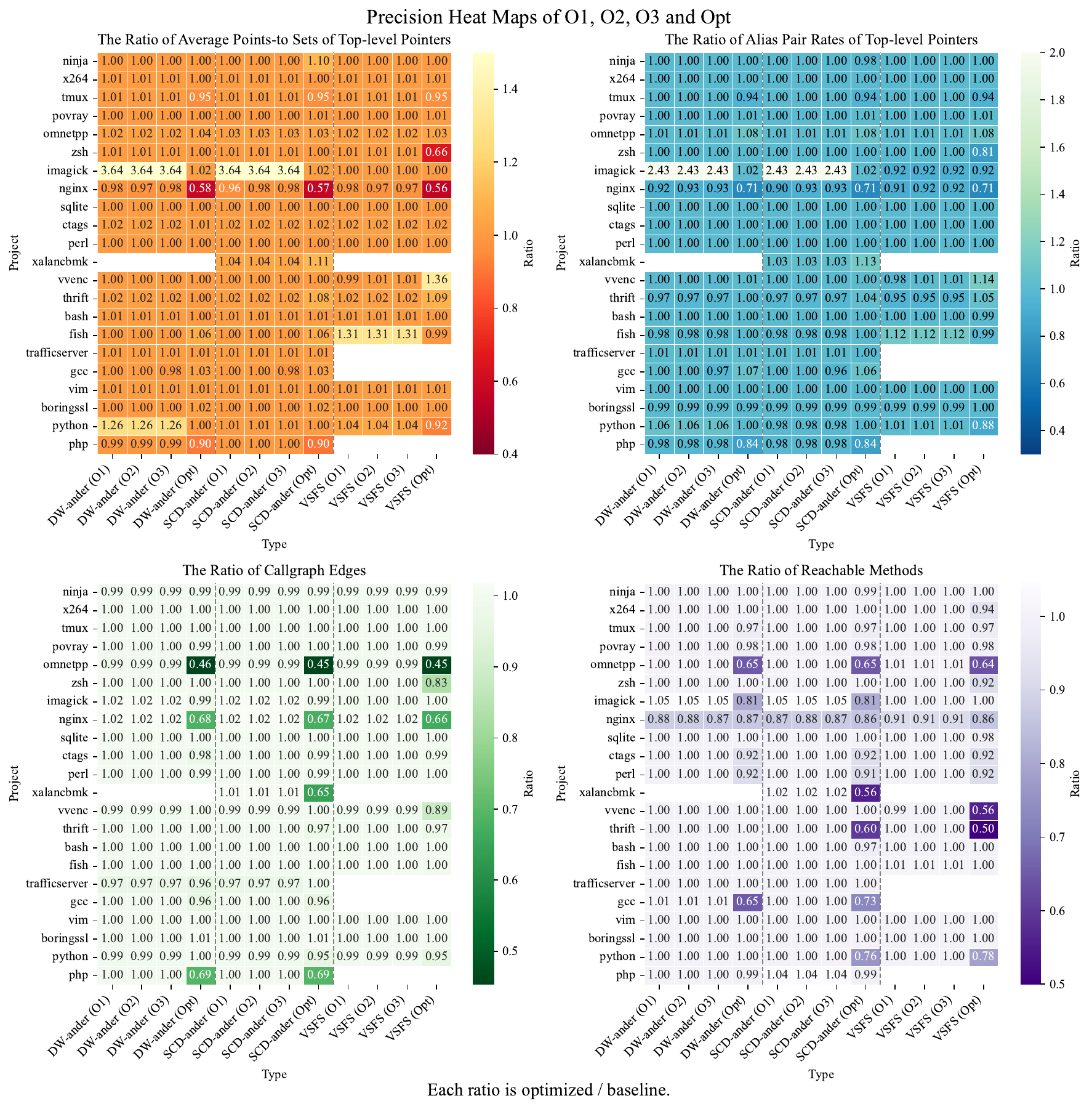}  % Replace with your PDF file
    \caption{Heat maps of precision change rate of O1, O2, O3, and \ToolName. Each ratio is optimized / baseline.}
    \label{fig:heatmap-precision}
   % \vspace{-3mm}
\end{figure}

%\yao{We can later decide whether to keep this part or not}
This section evaluates the impact of compiler optimizations on the precision of pointer analysis.
We employ four standard precision metrics, summarized in Table~\ref{tab:precision_metrics}.
Figure~\ref{fig:heatmap-precision} presents these metrics across a range of benchmarks and analysis configurations.
Key findings are outlined below.
%\yao{add remark that we do not use the alias analysis inside LLVM self}

\begin{table}[t]
\centering
\caption{Precision metrics for pointer analysis}
\label{tab:precision_metrics}
\begin{tabular}{|l|l|}
\hline
\textbf{Metric} & \textbf{Description} \\
\hline
Points-to sets & Average size of points-to sets of top-level pointers \\
\hline
Alias pairs & Ratio of identified aliased pairs of top-level pointers \\
\hline
Callgraph edges & Number of callgraph edges for function pointer resolution \\
\hline
Reachable methods & Number of functions reachable from entry point \\
\hline
\end{tabular}
%\vspace{-3mm}
\end{table}

\smallskip
\noindent \textbf{Points-to Set Size}.
For most benchmarks, the average points-to set size remains stable across optimization levels. However, several programs exhibit notable changes with \ToolName s applied. 
For example, in \texttt{nginx}, sizes decrease by $42.4\%$ under DW-ander, $43.1\%$ under SCD-ander, and $43.7\%$ under VSFS. \texttt{Php} sees reductions of $9.6\%$ under DW-ander and $10.2\%$ under SCD-ander. \texttt{Zsh} and \texttt{python} also show decreases of $34.4\%$ and $7.6\%$, respectively, under VSFS.
Conversely, some projects experience increased points-to set sizes, suggesting reduced precision. \texttt{Vvenc} grows by $35.6\%$ under VSFS, while \texttt{xalancbmk} increases by $10.8\%$ under SCD-ander. 
% These cases indicate that the optimization does not uniformly improve precision.

\smallskip
\noindent \textbf{Alias Pair Ratio}. 
%While most projects exhibit minimal changes, some experience significant improvements or degradations, indicating that the optimization has a limited overall effect but a notable impact in specific cases.
The aliasing ratio remains largely unchanged for most programs, but a few exhibit some variation.
In \texttt{nginx}, \texttt{php}, \texttt{zsh}, and \texttt{python}, the optimization reduces aliasing, improving precision. For \texttt{nginx}, the aliasing ratio decreases by $19.86\%$ under DW-ander, $19.88\%$ under SCD-ander, and $19.86\%$ under VSFS. \texttt{Php} sees reductions of $5.50\%$ under DW-ander and $5.54\%$ under SCD-ander. \texttt{Zsh} and \texttt{python} exhibit decreases of $7.39\%$ and $1.82\%$, respectively, under VSFS.
Conversely, \texttt{vvenc}, \texttt{xalancbmk}, and \texttt{omnetpp} show increased aliasing rates, suggesting reduced precision. 
 However, these degradations are relatively small.
 % compared to the improvements observed in other projects, indicating that the optimization generally enhances precision.

%\smallskip \emph{Callgraph Ddges}.\yao{Callgraph Ddges and Reachable methods, maybe just keep one} Most projects experienced only minor changes, with a few exceptions showing significant improvements in precision. 

\smallskip
\noindent \textbf{Callgraph Edges}.
Most benchmarks exhibit minor changes in call graph size, with a few showing substantial precision gains. Among the projects, \texttt{omnetpp} is a notable example where optimization results in substantial reductions in call graph edges. Under DW-ander, the edge count decreased from $258,683$ to $117,774$, a reduction of $54.5\%$. Similarly, under the SCD-ander and VSFS, the reductions are $54.5\%$ and $54.7\%$, respectively. Another notable example is \texttt{php}, where the edge count decreased by $30.7\%$ under both DW-ander and SCD-ander. Additionally, \texttt{xalancbmk} exhibits a significant reduction of $35.3\%$ under SCD-ander. \texttt{Nginx} also experiences evident reductions of $32.0\%$, $32.6\%$, and $33.5\%$, respectively. 
% These improvements highlight the effectiveness of optimization in reducing unnecessary edges and enhancing precision in these specific projects. 
In conclusion, while the optimization has a limited effect on most projects, it demonstrates the ability to reduce callgraph edges in specific cases. 

\smallskip
\noindent \textbf{Reachable Methods}.
 We measure the number of functions reachable from the entry point.
    % of a program.
Similarly, most benchmarks exhibit minimal changes, though some see notable improvements. \texttt{Omnetpp} shows a significant reduction in reachable functions, decreasing by $35.2\%$ under DW-ander and $36.4\%$ under VSFS. Similarly, \texttt{xalancbmk} drops by $44.5\%$ under SCD-ander and \texttt{vvenc} by $44.1\%$ under VSFS, demonstrating effective pruning of infeasible functions. Moderate improvements appear in \texttt{gcc} ($35.1\%$ under DW-ander, $27.3\%$ under SCD-ander), \texttt{python} ($21.5\%$ under VSFS), and \texttt{thrift} ($50.1\%$ under VSFS). In contrast, \texttt{boringssl} and \texttt{zsh} show slight increases, with negligible impact on precision. Overall, optimizations have a limited effect on most benchmarks but significantly improve precision in complex programs such as \texttt{omnetpp} and \texttt{xalancbmk}.

%\smallskip \noindent \textbf{Differences of Different Analyses}. \yao{Any diff among the three analyses?} Figure~\ref{fig:heatmap-precision} illustrates the impact of O1, O2, O3, and \ToolName\ on four precision metrics (as defined in Section~\cref{subsec:precision}). In general, precision remained largely unchanged under O1, O2, and O3. However, certain programs exhibited notable variations. For instance, \texttt{imagick} experienced a significant drop in points-to-set and alias pair precision under standard optimizations. In contrast, \ToolName\ improved precision for several programs, including \texttt{nginx}, \texttt{omnetpp}, \texttt{xalancbmk}, \texttt{gcc}, \texttt{python}, and \texttt{php}, across different metrics. 

%The results indicate that there was usually no variance in precision across different techniques. However, there were also exceptions. For \texttt{imagick}, both DW-ander and SCD-ander showed a sharp decrease in the precision of points-to sets, while VSFS remained stable. For \texttt{zsh} and \texttt{vvenc}, precision witnessed a notable increase in points-to sets and reachable methods, respectively, under VSFS.  We can conclude that precision remained stable in most cases, but if it changed, VSFS was usually different from DW-ander and SCD-ander.

\mybox{\finding\ Compiler optimizations typically have limited impact on pointer analysis precision, but can cause substantial improvements or degradations in specific cases. Their effect is non-monotonic and varies across programs and analyses, with some optimizations even reducing the precision metrics.}

In this work, our focus is on improving scalability through compiler-based offline simplifications.
Future work could explore strategies for balancing precision and efficiency and identify optimization passes that significantly impact precision. Understanding the precise effects of compiler optimizations on pointer analysis could lead to more effective and targeted optimization techniques. 
% \gu{Discarding passes which are not important from the waiting list to narrow the search space.}

\begin{remark} Compiler optimizations modify the IR, which may affect the comparability of analysis results across configurations. Nonetheless, the relative stability of the metrics across benchmarks and analyses provides evidence about the general trends in analysis precision. We acknowledge, however, that IR transformations introduce threats to validity that cannot be entirely ruled out.
\end{remark}

\section{Discussions}
\label{sec:discuss}
In this section, we discuss strategies to enhance the effectiveness and applicability of our approach, as well as potential threats to validity. 

%\yao{It is better to recall the findings/data in the evaluation instead of just ``brainstorming''...; highlight the implications of this work instead of discussing some high-level points.}

%The 300 combinations initially explored may not fully capture the compiler optimization space. We conducted an additional experiment to assess whether this number is sufficient, expanding the number of combinations to 8000. Due to time constraints, each combination was evaluated on a randomly selected subset of xxx. The optimal speedups achieved across these 8000 combinations are shown in the Figures. Compared with the results in Figure xxx, we observe no significant differences, except for a few outliers. This suggests that the 300 combinations may adequately cover the optimization space within a reasonable time frame. \yao{if we have enough time..} \yao{maybe different observations..}

\subsection{Improving the Effectiveness} 
\label{subsec:effectiveness}

Here, we discuss several potential avenues to enhance the effectiveness of our approach.

\smallskip
\noindent \textbf{Exploring the Optimization Space}.
We follow the iterative optimization approach in the compiler community, where a program is compiled multiple times using different compiler flags to select the best configuration.
There are several avenues to better explore the compilation space.
\begin{itemize}
    \item \emph{Guided Iterative Search}.
    We use a simple random search strategy to evaluate various pointer analyses while maintaining consistent compiler configurations.
    More sophisticated search techniques—such as Bayesian optimization, simulated annealing, or genetic algorithms—could accelerate convergence toward better configurations.
    %However, this process can be prohibitively time-consuming, requiring multiple recompilation steps to identify the optimal configuration.
    \item \emph{The Phase Ordering Problem}. A fundamental challenge in iterative optimization is phase ordering: determining the optimal sequence of optimization passes~\cite{staticrl}. Our study does not account for the ordering or repetition of passes, yet prior work suggests that phase ordering significantly impacts performance. Addressing this issue would expand the search space and could also unlock additional optimization opportunities.
    \item \emph{Fine-Grained Simplification Scope}. We restrict our focus to ``one-configuration-fits-all'' optimizations, where a single combination of compiler optimizations is applied uniformly across the entire program. 
     A more refined approach—where different optimizations are applied to specific program regions (e.g., individual functions)—could improve performance. Notably, the effect of an optimization on one function may influence the optimal configuration for another, introducing complex interdependencies that merit further investigation.
\end{itemize}

\smallskip
\noindent \textbf{Annotations for Guiding Transformations}. 
Our evaluation shows that simplification passes can substantially influence the precision and scalability of pointer analysis. While many transformations can be applied automatically, developers and analysis designers may benefit from mechanisms that guide these transformations more effectively.

Lightweight source-level annotations or compiler hints (e.g., \texttt{\_\_attribute\_\_((aligned))}, \texttt{restrict}, or custom pragmas) could provide a more systematic approach to guiding simplification. By explicitly marking optimization-relevant properties—such as pointer aliasing constraints, alignment requirements, or intended access patterns—developers could direct transformations toward analysis-friendly forms without exhaustive search. 

While manual annotation imposes an additional burden, semi-automated inference techniques (e.g., based on program analysis or machine learning) could reduce this overhead. Such annotation-guided optimization would complement our approach, offering a hybrid strategy that balances automation with developer expertise.

\smallskip
\noindent \textbf{Predicting Optimization Strategies}. 
An alternative to iterative search is predictive modeling, where optimization configurations are selected based on static or dynamic program features in conjunction with the target analysis. Recent work has demonstrated the viability of machine learning techniques in guiding compiler optimizations, including vectorization~\cite{mendis2019ImitationLearning} and function inlining~\cite{trofin20MLGO}. These approaches aim to bypass expensive search by learning heuristics from prior optimization outcomes.
However, predictive models often struggle to match the performance of exhaustive or adaptive search strategies in compiler auto-tuning~\cite{Cooper-Adaptive21stCentury}. This gap is largely due to the difficulty of capturing complex, non-linear interactions among transformations, which are highly sensitive to both program structure and optimization context.

Whether similar limitations arise in the context of pointer analysis—particularly when the goal is to improve precision or reduce analysis cost—remains an open question.
Moreover, the effectiveness of predictive models may depend on the granularity of the prediction task: coarse-grained strategies (e.g., selecting a global transformation sequence) may be less effective than fine-grained ones (e.g., deciding whether to apply a specific simplification at a given program location). Exploring this tradeoff, and identifying features that correlate with analysis outcomes, is a promising direction for future work.
% Investigating this direction could determine whether predictive models can effectively identify optimization configurations that are either analysis-agnostic or specialized for particular analyses.

\subsection{Extending the Applicability}
\label{subsec:applicability}
Compiler-based offline simplification for pointer analysis could be utilized in broader contexts.
%offering potential improvements in other contexts.

\smallskip
\noindent \textbf{Studying other Pointer Analyses}. 
While our study focuses on several exhaustive points-to analyses,
our evaluation framework could be adapted to other pointer analyses. 
\begin{itemize}
    \item \emph{Beyond the Exhaustive Analyses}. In addition to the exhaustive analysis, other types of analyses have been studied.
First, the demand-driven analysis (for C/C++~\cite{heintze2001demand,zheng2008demand} and Java~\cite{sridharan2005demand,sridharan2006refinement,yan2011demand,shang2012demand}) only analyzes the program's relevant parts for answering a given query.
%wei2015adaptive
% saha2005incremental
%lu2013incremental
%kastrinis2013hybrid,
% zhang2014abstraction,
Second, \emph{selective x-sensitive analysis}~\cite{smaragdakis2014introspective,jeon2018precise,li2018scalability} resembles demand-driven analyses. However, they are not query-driven but schedule analysis strategies, such as selective context sensitivity for different program elements. 
Compiler optimizations may alter the metrics used by previous work, and our study can help inform the design of optimization-resistance experiments.
%evaluate the compiler-agnostic capability
Third, the \emph{iterative refinement} approach~\cite{yan2024scaling} uses progressively more precise analyses to verify specific queries, enabling gradual precision improvements while balancing overhead.
One avenue for future research could be leveraging program transformation as a form of refinement, as in control-flow refinement~\cite{Gulwani2009,Balakrishnan2009,Flores-Montoya2014}.
%to enhance analysis precision in the context of program verification.
    \item  \emph{Store-based vs. Storeless Abstractions}. 
We have concentrated on points-to analysis, which, for each program entity (such as a variable or object reference), determines the set of memory locations that may be indirectly referenced through that entity. However, other pointer analyses mainly use 
a storeless abstraction~\cite{kanvar2016heap}, such as tracking aliased access paths~\cite{hackett2006aliasing}, computing alias sets~\cite{naeem2009efficient}, and more. A promising direction for future work is to extend our evaluation to encompass these broader forms of analysis.
%\yao{a big longer}
\end{itemize}

\noindent \textbf{Improving IR Lifted from Binaries}. 
Binary lifting translates low-level executables into higher-level intermediate representations (IR) to enable static analysis and reverse engineering~\cite{kvroustek2017retdec,zhou2024plankton,dinaburg2014mcsema}. The lifted IR may adopt custom formats or reuse existing infrastructures such as LLVM IR. However, the lifting process is inherently lossy: compilation strips away high-level abstractions, including types, control-flow structure, and variable naming, making accurate reconstruction difficult.

A recent study~\cite{liu2022sok} evaluated four state-of-the-art binary lifters targeting LLVM IR and found that the resulting IR often lacked sufficient semantic fidelity to support precise pointer analysis. In particular, the absence of type information—especially for pointers—severely limited the effectiveness of downstream analyses. As a result, the lifted IR frequently failed to yield actionable insights in practical settings.
Recent efforts have explored applying standard compiler optimizations to lifted IR~\cite{zhou2024plankton,retdec}, aiming to improve its structure and analyzability. A promising direction is to specialize these optimizations to explicitly support pointer analysis. For example, transformations that expose memory access patterns, eliminate spurious control flow, or infer likely type information could significantly improve analysis precision. Tailoring the optimization pipeline to the characteristics of lifted code may thus offer a path toward more effective binary analysis.
%Binary lifting translates low-level binary executables into higher-level IR to facilitate program analysis and comprehension~\cite{kvroustek2017retdec,zhou2024plankton,dinaburg2014mcsema}. The IR may be custom-built or based on established formats, such as LLVM IR. However, lifting binaries poses significant challenges, primarily because critical information is lost during compilation. A previous study~\cite{liu2022sok} evaluated four state-of-the-art binary lifters that generate LLVM IR, revealing substantial limitations in pointer analysis. The lifted IR often produced outputs with limited practical value, primarily because it was difficult to recover type information—especially for pointers—from stripped binaries.  Recent works have used compiler optimizations for lifted LLVM IR~\cite{zhou2024plankton,retdec}. An interesting direction is to tailor compiler optimizations to enhance pointer analysis, e.g., to facilitate better recovery of type information. 

\smallskip \noindent \textbf{Validating Pointer Analyses}. We examine the impact of compiler-based offline simplifications on pointer analysis. However, the interaction between program transformations and static analysis can introduce flaws that are difficult to detect and debug~\cite{Do2020TSEDebugSA}. First, LLVM’s optimizer is not immune to bugs, and applying optimization passes in unexpected orders may expose subtle correctness issues. In the compiler testing community, several efforts have utilized combinations of compiler flags~\cite{chen2022boosting,jia2023detecting}. Second, even if the compiler is free of bugs, the mutated IR may reveal flaws in the analysis tools, as they may make specific assumptions about the IR's structure or fail to support specific instructions. Indeed, as a byproduct, \emph{when running the experiments, we found twelve issues in SVF. We reported them to the developers, and several have been fixed.}

%%%%%%%%%%%

%PhASAR~\cite{schubert2019phasar} proposes a methodology for debugging static analyzers in the presence of optimizations. The process begins by exhaustively debugging the static analyzer on test cases using unoptimized IR. Optimizations are incrementally enabled once the analyzer is verified in this baseline setting. Differential testing is then performed, using the unoptimized IR results as a reference. \yao{to revise}

%\cite{Do2020TSEDebugSA} identified handling corner cases involving complex IR constructs as one of the most common sources of errors when debugging analysis tools. 

%\yao{debug info validation by Qirun Zhang, etc?}
%\cite{zhang2023statfier}

%\smallskip \noindent \textbf{Improving the Precision?}. \yao{maybe discuss briefly here. If the readers ask for xx, we can put the extended content in the previous sections.}

\subsection{Threats to Validity}
\label{subsec:validity}
Our benchmark suite may not fully capture the diversity of possible program behaviors. Thus, we cannot claim that our conclusions generalize beyond the benchmarks of this study. Nevertheless, the results exhibit consistent trends across all evaluated benchmarks, suggesting that the observed patterns may extend to a broader class of programs.
Besides,  while our current evaluation focuses on specific precision metrics, a more comprehensive assessment would involve additional client applications, such as program verification and vulnerability detection. However, using these high-level clients can be problematic from a methodological perspective, as they often depend on more than just an accurate heap abstraction (e.g., reasoning numerical values and path conditions). These additional dependencies may introduce further sources of variability.

 \section{Related Work}
\label{sec:related}

%\yao{There is a work studying the impact of IR for Java pointer analysis, which we should discuss}

\noindent \textbf{Staged Approach to Pointer Analysis}. 
Our method follows the staged approach to pointer analysis, which uses auxiliary analyses to bootstrap the subsequent primary analysis, such as \emph{simplifications}~\cite{Rountev2000,hardekopf2007exploiting,Li2020,Lei2023}, \emph{sparsification}~\cite{yan2018spatio}, \emph{pruning}~\cite{fink:typestate:issta}, \emph{partitioning}~\cite{kahlon2008bootstrapping}, and \emph{indexing}~\cite{shi2022indexing}.
Existing offline simplifications are typically tailored to the analyses they support.
\citet{hardekopf2007exploiting} uses pointer and location equivalence among variables in the Andersen analysis, facilitating the consolidation of equivalent nodes and subsequent graph simplification. 
\citet{Li2020} proposes an approach that eliminates graph edges irrelevant to InterDyck-paths. 
\citet{Lei2023} introduce criteria for identifying foldable node pairs, leveraging an RSM-based graph-folding algorithm.
Sparse analysis utilizes an auxiliary analysis to pre-compute def-use information~\cite{hardekopf2011flow,sui2016sparse,sui2016demand,yan2018spatio}, thereby reducing the redundant propagation of facts in the primary analysis. 
%A more accurate auxiliary analysis in the staged approach allows better sparsification, fine-grained partitions, and improved overall disambiguation. 
However, they must balance the effectiveness and overhead of the auxiliary analysis~\cite{hardekopf2011flow,shi2018pinpoint}: a precise one limits scalability, while an imprecise one seriously undermines precision.
An intriguing aspect of our approach is that more aggressive compiler optimizations do not necessarily yield proportional performance improvements in subsequent analyses. 
The relationship between optimization aggressiveness and analysis efficiency is non-trivial and needs further investigation.

%\yao{more systematic/in-depth discussion? And discuss our work?}

\smallskip
\noindent \textbf{Code Transformations for Static Analysis}.
%\yao{TBD: improve this part}
\citet{cousot2002systematic} develop a systematic theory of semantically justified program transformations based on abstract interpretation. Their framework provides a formal foundation for proving that a transformed program preserves specific semantic properties of the original program. 
Program transformations have been widely explored to improve the precision of static analysis, particularly through \emph{control-flow refinement} \cite{Gulwani2009,Balakrishnan2009}, which enhances the control-flow graph to distinguish execution paths more precisely. Examples include predicate splitting~\cite{Sharma2011} and loop rewriting~\cite{Gulwani2009}. 
Other transformation approaches are ``implicit'', focusing on CFG elaboration but primarily enabling efficient computation over a refined powerset extension of the program's state space, such as \emph{trace partition}~\cite{rival2007trace} and \emph{elaboration} \cite{Sankaranarayanan2006}.
A related line of work examines the interaction between compiler optimizations and abstract interpretation~\cite{namjoshi2018impact}.
%present a formal framework for analyzing how compiler optimizations affect the precision of numerical abstract domains. 
In contrast, our focus is on improving scalability through program transformations. %\citet{prakash2021effects} study the impact of IR on Java pointer analysis frameworks, but do not explore compiler-based transformations to improve analysis.
% \yao{should discuss this paper better}

\smallskip
\noindent \textbf{Iterative Compiler Auto-Tuning}.
Modern compilers offer a wide range of optimization passes, whose interactions are often complex and difficult to predict. 
Auto-tuning has been extensively studied in the compiler community to automate the selection of effective optimization sequences. Most efforts focus on the \emph{iterative optimization}~\cite{supersonic,ansel2014opentuner} paradigm, which recompiles the program multiple times with different compiler configurations. 
Various search algorithms have been explored, such as Bayesian optimization~\cite{chen2021efficient} and genetic algorithms~\cite{2003Stephenson}, utilizing feedback such as assembly instruction count, binary code size, and runtime.
\citet{2003Stephenson} apply genetic algorithms to tune heuristic priority functions for three compiler optimization passes.  penTuner~\cite{ansel2014opentuner} autotunes a program using an AUC-Bandit-meta-technique-directed ensemble selection of algorithms.
%, which does not consider the order or support repeated optimizations.  
Recently, \citet{chen2021efficient} proposed BOCA, a Bayesian optimization approach that leverages a tree-based model and a novel strategy to efficiently tune compiler flags.
Machine learning techniques have been employed to guide the iterative selection process based on the program's characteristics~\cite{staticrl, supersonic,10.1145/3480250}, which enables the evaluation of a range of possible options without compiling and profiling each one.
The models are used to predict the speedups of candidate sequences~\cite{Park2012}, determine the superior sequence between two candidates~\cite{Purini2013}, or directly suggest candidate sequences~\cite{Park2011}.
% \citet{ren2021unleashing} uses auto-tuning to study the impact of optimizations on binary similarity analysis.
We focus on offline simplifications to speed up pointer analysis. 

%in~\cite{2012Kulkarni}, the challenge of pass selection is formulated as a Markov process, and supervised learning was used to predict the subsequent optimization based on the current program state.

%A speedup predictor takes as input both the characterization of the program being compiled and an optimization sequence, and it predicts as output the speedup when applying that optimization sequence relative to a default optimization setting.

%A sequence predictor characterizes a program being compiled and uses it as input to a model, and the model predicts a probability distribution of optimizations to apply to that program.

%A tournament predictor takes as input a triple corresponding to the characterization of the program and two optimization sequences. This model predicts whether the speedup achieved after applying the first optimization sequence will be greater or less than the speedup obtained with the second optimization sequence.

%\citet{staticrl, supersonic} employs reinforcement learning for the phase-ordering problem. 
\section{Conclusion}
\label{sec:conclu}
%This paper introduces an analysis-agnostic approach to offline simplification in pointer analysis by repurposing standard compiler optimizations. 
% (e.g., to create specific code components)
We have presented a new perspective on offline simplification for pointer analysis. Unlike prior techniques, the approach is analysis-agnostic, enables diverse simplification opportunities, and facilitates instance-specific simplifications. On the practical side, it supports modular and selective applications, is readily applicable to existing analyses, and composes naturally with prior simplifications.
Our empirical evaluation demonstrates that the approach improves end-to-end performance, alters key IR metrics, and influences the precision of analysis. We also characterize the space of optimization configurations and their interaction with downstream analyses. These results suggest new directions for adaptive simplification strategies and optimization pipelines tailored to the needs of pointer analysis.
%We conduct an empirical study that yields several key findings, including improvements in end-to-end performance, the effects on IR metrics, the characteristics of optimization pass configurations, and the impact of optimizations on analysis precision. Our work also opens avenues for further research, such as adaptive optimization, selection, and transformations specialized for pointer analysis.
%  while maintaining precision. 

%However, balancing transformation complexity with resulting performance gains remains a key challenge.

%Despite advances in offline optimization for pointer analysis, prior work has been typically tailored to individual analyses. In this paper, we formulate the optimization as a preprocessing step that is largely orthogonal to subsequent pointer analysis. To demonstrate the generality of our approach, we applied it to three distinct pointer analysis algorithms and two analysis frameworks. 

\section{Data Availability}
\label{sec:data}

The tool and dataset are available at \url{https://anonymous.4open.science/r/opt4pta-D464/}.
%% Bibliography style
\bibliographystyle{ACM-Reference-Format}
%\citestyle{acmauthoryear}   %% For author/year citations
\bibliography{alias,compiler,tmp}

\end{document}